\RequirePackage{fix-cm}
\documentclass[twocolumn]{svjour3}          
\smartqed  
\usepackage{algorithm}
\usepackage{algorithmic}
\usepackage{multirow}
\usepackage{graphicx}
\usepackage{subfigure}

\begin{document}

\title{Learning Discriminative Representations for Semantic Cross Media Retrieval
}
\subtitle{}


\author{Aiwen Jiang         \and
        Hanxi Li	\and
        Yi Li	\and
        Mingwen Wang
}


\institute{Aiwen Jiang \at
              NICTA, London Circuit 7, ACT, Australia \\
              \email{Aiwen.Jiang@nicta.com.au}           
           \and
           Yi Li \at
              NICTA, London Circuit 7, ACT, Australia \\
           \and
           Hanxi Li, Mingwen Wang \at Jiangxi Normal University, China
}

\date{Received: date / Accepted: date}

\maketitle

\begin{abstract}
Heterogeneous gap among different modalities emerges as one of the critical issues in modern AI problems.
Unlike traditional uni-modal cases, where raw features are extracted and directly measured, the heterogeneous nature of cross modal tasks requires the intrinsic semantic representation to be compared in a unified framework.
This paper studies the learning of different representations that can be retrieved across different modality contents.
A novel approach for mining cross-modal representations is proposed by incorporating explicit linear semantic projecting in Hilbert space.
The insight is that the discriminative structures of different modality data can be linearly represented in appropriate high dimension Hilbert spaces, where linear operations can be used to approximate nonlinear decisions in the original spaces.
As a result, an efficient linear semantic down mapping is jointly learned for multimodal data, leading to a common space where they can be compared.
The mechanism of "feature up-lifting and down-projecting" works seamlessly as a whole, which accomplishes crossmodal retrieval tasks very well.
The proposed method, named as shared discriminative semantic representation learning (\textbf{SDSRL}), is tested on two public multimodal dataset for both within- and inter- modal retrieval.
The experiments demonstrate that it outperforms several state-of-the-art methods in most scenarios.
\end{abstract}

\section{Introduction}\label{sec:introduction}
Cross-media or crossmodal retrieval emerges as an important research area that attracts great interests from artificial intelligence, multimedia information retrieval and computer vision communities.
Unlike traditional unimodal tasks, where queries are in the same domain as the samples in repositories,
cross-media retrieval systems aim at understanding the matches that have similar semantic meanings but across different modalities.
For example, in the task of trying to search semantically relevant textual documents for a given visual data, or vice verse.
This mechanism makes the retrieval system much closer to our human behaviors than simple within-domain matching.

The difficulties of retrieval across different modalities lie in the fact that data from different domains have different representations.
Beyond the semantic gap from low level raw features to high-level semantics within domain,
there also exists heterogeneous-gap across domains that raw features from different modalities have different dimensions and physical meanings.
These facts make directly measuring similarities between different modality data very challenging.
Therefore, much effort has been made on mining the correlations and representations for multimodal data.

This paper aims to learn effective representations that can map different modality data into a common latent semantic space, where data from different modalities are measurable and retrievable.
We have proposed a mechanism with two stages working together to achieve this goal.
First, through kernel theory, raw data are non-linearly lifted up to an appropriate high dimension Hilbert space, where the intrinsic discriminating structures of data are preserved.
Then, we learn a multimodal based linear down projecting strategy to effectively mapping data from different modalities into intrinsic, shared semantic space.
The mechanism works seamlessly as a whole, which can accomplish crossmodal retrieval tasks very well.
We name it as Shared Discriminative Semantic Representation Learning (\textbf{SDSRL}).

The motivation on applying linear semantic down-projecting in Hilbert space is straightforward.
Our goal is to efficiently obtain data $X$'s semantic representation $Z_X$ through learning a linear down projecting matrix $A$, i.e., $Z_X = X*A$.
The projections $A$ is treated as linear discriminating hyperplanes. In this case, it requires the data should have linear discriminative structures 
thus the semantic space can be spanned by its intrinsic latent feature vectors. However, in most cases, handcrafted raw features generally lie in nonlinear separable manifolds.
In addition, in some special domains, representations are not always in the form of vectors. For example, manifold representations prefer structural descriptors such as co-variance matrix \cite{covariance_descriptor:eccv2006}.
It is unsuitable to directly applying linear projections on them. 

Nevertheless, according to the kernel theory, mapping descriptors to a Reproducing Kernel Hilbert Space (RKHS) is a promising strategy.
After having RKHS mappings, we can naturally design multimodal based algorithms in linear spaces for cross modal tasks.
Beside frequently used kernels, several valid kernel functions have recently been defined for non-vector data \cite{harandi2014eccv}.
Therefore, embedding into Hilbert space allows our proposed method having much more potentials to generalize well with more modality cases.

The main difficulty arises from the fact that Hilbert space may be infinite-dimensional, and the mapping to Hilbert space corresponding to a given kernel is also typically unknown. 
In order to overcome these problems, we adopt a nonlinear kernel approximation strategy\cite{fastfood_icml2013}\cite{SiSI_ICML2014}\cite{Nystrom_JMLR2005} to obtain an explicit mapping to Hilbert space. 
As a result, it preserves the advantages of orginal kernels, while enabling data having representations of finite dimension vector forms. 
Furthermore, the approximated mappings also facilitate our linear-fashion multimodal semantic projections learning in the later stage.

We have conducted experiments on two widely accepted multimodal dataset: Wikipedia dataset \cite{Xmodal_TPAMI2014} and NUSWIDE \cite{nuswide-civr09}. Both within and inter modal retrieval tasks are considered. The evaluation criteria are Mean Average Precision and Precision-Recall Curves. The experiment results show that our proposed method can not only do cross-modal retrieval with high performance, but also retain its advantages on unimodal retrieval tasks.

The contributions of this paper are twofold:
(1) We have proposed a flexible framework that seamlessly incorporates explicit linear semantic projecting in Hilbert space for mining cross-modal representations.
The generalization ability empowers our method to outperform many other state-of-the-art methods in cross-modal retrieval tasks.
(2) An efficient multimodal semantic projection learning strategy has been proposed by explicit modeling semantic correlations both within and between modalities. 
Compared with other strategies, it has much fewer parameters to be tuned, and less computational complexity. It also has much potentials to conveniently access different modality contents in embedding space for cross-modal tasks.

\section{Related work}

A large part of existing work on cross media retrieval focuses on mining correlations and representations for multimodal data.

One typical category of methods is based on Latent Dirichlet Allocation related topic models, such as Correspondence LDA \cite{CorLDA2003}, and Multimodal Document Random Field (MDRF) \cite{Jiayanqing_ICCV2011}.
The Correspondence LDA, extends Latent Dirichlet Allocation to find a topic-level relationship between images and text annotations,
in which distributions of topics act as a middle semantic layer. The MDRF suggests a mixture probabilistic graphical model using a Markov random field over LDA.
These methods are conducted in an unsupervised manner.

Some other work, like Semantic Correlative Matching \cite{Xmodal_TPAMI2014} and Multimodal Deep Learning \cite{Ngiam_ICML2011},
build lower-dimensional common representations based on canonical correlation analysis (CCA).
Menon \cite{menon-SDM2015} et al. proposed to reduce cross-modal retrieval as binary classification over pairs. The framework can be seen as a variant of Semantic Matching.
It is not difficult to find that, these approaches usually exploit the symbiosis of multimodal data in the assumption of there existing one-to-one strictly paired training data.

There are also a great amount of work focusing on hamming space as the semantic space. Multimodal hashing methods in this category includes
 Inter Media Hashing (IMH) \cite{IMH:2013SIGMOD}, Latent Semantic Sparse Hashing (LSSH) \cite{ZhouJile2014_LSSH}, CMFH\cite{CMFH_2014}, CSLP\cite{ZhenAAAI15}, QCH \cite{WuYZ_IJCAI15} and 
 Regularized CrossModal Hashing (RCMH) \cite{Moran-SIGIR2015} etc.
The IMH adopts a linear regression model with regularization to learn view-specific hash functions. The LSSH applies sparse coding and matrix factorization to learn the latent structure of image and text, 
then maps them to a joint hamming space. QCH takes both hashing function learning and quantization of hash codes into consideration. RCMH is an extension of Graph Regularized Hashing used in unimodal case.
Despite of the efficiency on retrieval time and memory storage brought by hashing, directly projecting raw features into short binary codes is in some extent an ``over compressing'' idea,
which may result severe reduction in retrieval accuracy.

In order to account for both the retrieval efficiency and accuracy, supervised common latent semantic features are learned in many recently published literature.
Typically, The CMSRM \cite{Wufei2013} considers learning a latent cross-media representation from the perspective of a listwise ranking problem.
The Multimodal Stacked AutoEncoders (MSAE) \cite{SAE_VLDB2014}, constructs stacked auto-encoders (SAE) to project features of different modality into a common latent space.
The Correspondence Autoencoder (CorrAE) \cite{wangxiaojie_MM2014} incorporates representation learning and correlation learning into a single process.
The  Bilateral Image-Text Retrieval (BITR) \cite{im2text_VermaJ14} addresses the problem of learning bilateral associations between visual and textual data based on structural SVM.
The data driven semantic embedding proposed by Habibian et al. \cite{HabibianICMR2015} is trained for image and text independently.
It should be noted that some of these supervisory methods have complex learning process with much parameters to be tuned, or they are unable to conveniently access different modality independently in learned mutual space. We are of the opinion that a simple and effective method is more preferred.

\section{The Proposed method\label{proposedmethod}}

We consider two modalities, image and text, in this paper.
Denote $X = \{ X_1 ,...,X_{n_1} \}$ for image documents and $Y = \{ Y_1 ,...,Y_{n_2} \}$ for text.
$n_1$ and $n_2$ are data size of image set and text set.

Firstly, we explicitly perform nonlinear feature lifting to map raw features $X$ and $Y$ to an appropriate high dimension feature space: $\Phi \left( X \right)$ for image data and $\Psi \left( Y \right)$ for text data.
Then, in the semantic down projecting stage, data from different modality is projected into a metric-comparable and semantic-embedded feature space, through two learned down projecting functions $A$ and $B$. The final multimodal semantic features are represented in the form of:
\begin{equation}
 {\rm Z}_{X}  = \Phi  * A, ~~~~  Z_{Y}  = \Psi  * B
\end{equation}

\subsection{Explicit nonlinear feature lifting}
According to the Moore-Aronszajn Theorem, a symmetric, positive definite kernel $K(\cdotp,\cdotp)$ defines a unique reproducing kernel Hilbert space on orginal space $X$, denoted hereafter by $H$, with the property that
there exists a mapping $\phi: X \rightarrow H$, such that $ K(x_1,x_2)=\langle \phi(x_1), \phi(x_2) \rangle _H = \phi(x_1)^\prime \phi(x_2)$.

The main difficulty arises from the fact that $H$ may be infinite-dimensional, and, more importantly, that the mapping $\phi$ corresponding to a given
kernel $K$ is typically unknown. Kernel approximation is a recent trend, whose goal is to approximate a given kernel $K$ by a feature map that is finite dimensional.
\begin{equation}
 K\left( {x,y} \right) \approx \left\langle {\hat \Phi \left( x \right),\hat \Phi \left( y \right)} \right\rangle
\end{equation}
Applying a linear classifier on the resulted feature mappings $\hat \Phi \left( \cdotp \right)$ can yield non-linear decisions in the original space.

The original intention of kernel approximation is proposed as an effective solution to speed up the computation efficiency for nonlinear kernel methods, especially when dealing with large amount of data \cite{Corinna_AISTATS2010} \cite{VedaldiCVPR12} \cite{fastfood_icml2013}.
In this paper, from a view of feature representation, we propose to treat it as a practical feature transformation way, rather than simply a computation cost reduction method.

We adopt Nystroem approximation method \cite{Nystrom_JMLR2005} for its simplicity.
The basic idea is that, given a collection of M training examples $\{x_i\}_{i=1} ^M$  and their corresponding kernel matrix $K$, $i.e. [K]_{ij} = k(x_i,x_j)$,
we aim to find a matrix $Z \in R ^{r \times M}$, such that $K=Z^\prime Z$. The best approximation in the least-squares sense is given
by $Z =\Sigma^{1/2} V$ , with $\Sigma$ and $V$ the top $r$ eigenvalues and corresponding eigenvectors of $K$.
As a result, for a new sample $x \in X$, the $r$-dimensional vector representation $z(x)$ of the space induced by $k(x,\cdotp)$ can be written as
\begin{equation}
 z(x) =\Sigma^{1/2} V \left[ k(x,x_1), k(x,x_2), ..., k(x,x_M) \right]^\prime
\end{equation}

From Nystroem method, we obtain corresponding up-lifted feature mappings $\Phi \left( X \right) \in \Re ^{n_1  \times m_1 }$ for image and $\Psi \left( Y \right) \in \Re ^{n_2  \times m_2 }$ for text.
$n_1, n_2$ are dataset size of image and text. $m_1, m_2$ are dimensions of approximated feature mappings.

\subsection{Linear down projection for semantic retrieval}
The goal of this stage is to learn explicitly linear semantic down projecting matrices $A \in \Re ^{m_1 \times q}$, $B \in \Re ^{m_2 \times q}$ that can map different modality data into a low-dimensional semantic space. $q$ is the dimension length.

Formally, at the beginning, we define a semantic correlation
$S\left( {a,b} \right) = \sum\limits_j {\frac{2{L_a ^j L_b ^j }}{{L_a ^j  + L_b ^j }}}$
from supervised labels,
where the $j^{th}$ element of $L_l ^j (l=a , b)$ is set to be 1, if data \emph{l} has the $j^{th}$ semantic label, otherwise 0. This is motivated by the observations that data are often multi-labeled. With the definition, we can rank the semantic correlations among data.

Denoting homogeneous $S_I  \in \Re ^{n_{1}  \times n_{1}}$ for intra-image, $S_T  \in \Re ^{n_{2}  \times n_{2}}$ for intra-text, and heterogeneous $S_C \in \Re ^{n_{1}  \times n_{2}}$ for inter-image\&text. The final features $Z_X, Z_Y$ should satisfy optimization problem (\ref{opt_all}). Here, we consider a linear similarity function.
\begin{tiny}
\begin{equation}
\label{opt_all}
\mathop {\min }\limits_{A,B} \underbrace {\| {{S_I} - \left\langle {Z_X,Z_X} \right\rangle } \|_F^2}_{{\rm{intra-img}}} + \underbrace {\| {{S_T} - \left\langle {Z_Y,Z_Y} \right\rangle } \|_F^2}_{{\rm{intra-text}}} + \underbrace {\| {{S_C} - \left\langle {Z_X,Z_Y} \right\rangle } \|_F^2}_{{\rm{inter-img\& text}}}
\end{equation}
\end{tiny}
It is challenging to directly solve the whole optimization problem of (\ref{opt_all}) and obtain A and B. However, after having the up-lifted mappings $\Phi$ and $\Psi$, we can relax the difficulties by resolving (\ref{opt_all}) into two continuous sub-processes logically.

Firstly, we define $M_I = AA' \in {\Re ^{ m_1 \times m_1}}$, $M_T = BB' \in {\Re ^ { m_2 \times m_2}}$ as intra-modal semantic-link weight matrixes and $M_C = AB' \in {\Re ^{m_1 \times m_2}}$ inter-modal semantic-link weight matrix for multimodal data. If we take these $M$ as intermediate solutions, the problem (\ref{opt_all}) can be divided into three sub-problems: $\min \left\| {{S_I} - \Phi {{\rm{M}}_I}\Phi ^{'}} \right\|$, $\min \left\| {{S_T} - \Psi {{\rm{M}}_T}\Psi ^{'}} \right\|$, $\min \left\| {{S_C} - \Phi {{\rm{M}}_C}\Psi ^{'}} \right\|$, which can be optimized independently. The solutions $M_I ^*$, $M_T ^*$, $M_C ^*$ satisfy Problem (\ref{opt_all}).
\begin{eqnarray}
\begin{array}{c}
M_I ^*  = \left( {\Phi ^{'} \Phi  + \mu I} \right)^{ - 1} \Phi ^{'} S_I \Phi \left( {\Phi ^{'} \Phi  + \mu I} \right)^{ - 1}\\
M_T ^*  = \left( {\Psi ^{'} \Psi  + \mu I} \right)^{ - 1} \Psi ^{'} S_T \Psi \left( {\Psi ^{'} \Psi  + \mu I} \right)^{ - 1}\\
M_C ^* = \left( {\Phi ^{'} \Phi  + \mu I} \right)^{ - 1} \Phi ^{'} S_C \Psi \left( {\Psi ^{'} \Psi  + \mu I} \right)^{ - 1}
\end{array}
\end{eqnarray}
$\mu$ is regularized parameters with relative small value.

Then, we jointly learn projections $A$ and $B$ by minimizing loss (~\ref{totalloss}).
\begin{equation}
\label{totalloss}
Loss = \min_{A,B} \left\| {M_I ^*  - A A ^{'} } \right\| + \left\| {M_T ^*  - B B ^{'} } \right\| + \left\| {M_C ^*  - A B ^{'} } \right\|
\end{equation}
The resulted optimal $A \in \Re ^{m_1 \times q}$ and $B \in \Re ^{m_2 \times q}$, in its ideal case, consequently can satisfy the semantic correlations previously defined in (\ref{opt_all}).

The advantage of reducing problem from (\ref{opt_all}) to (\ref{totalloss}) is that the problem scale of (\ref{totalloss}) is irrelevant with specific task. It is computational affordable, even for large scale tasks, because $m_1$ and $m_2$, the dimension lengths of embedded mappings, are moderately smaller compared with dataset size $n_1, n_2$.

We minimize (\ref{totalloss}) by solving sub-problems (\ref{subop1}) and (\ref{subop2}) iteratively.
\begin{itemize}
 \item
 \textbf{Step1:} By fixing \textbf{B}, Optimize the Loss
  \begin{equation}
  \label{subop1}
  \mathop {\min }\limits_{A }  \left\| {M_I  - A A ^{'} } \right\| + \left\| {M_C  - A B ^{'} } \right\|
  \end{equation}

 \item
 \textbf{Step2:} By fixing \textbf{A}, Optimization the Loss
  \begin{equation}
  \label{subop2}
  \mathop {\min }\limits_{B } \left\| {M_T  - B B ^{'} } \right\| + \left\| {M_C  - A B ^{'} } \right\|
  \end{equation}
\end{itemize}
As the problem (\ref{subop1}) and (\ref{subop2}) are similar in form, they can be solved in a similar way.

The solution for every sub-problem (\ref{subop1}), named as MPL-CD, is shown in Algorithm \ref{alg:CD_MMSP}.
We repeat the learning process (\ref{subop1}) and (\ref{subop2}) for \emph{E} loops.

\begin{algorithm}
\caption{Coordinate Descent based Multimodal Projecting Learning (MPL-CD)}
\label{alg:CD_MMSP}
\begin{algorithmic}[1] 
\REQUIRE ~~\\ 
$M_I,~~M_C,~~\textbf{B}$.
Objective dimension \emph{q}; tolerance error $\varepsilon$; Maximum iterations T;
\ENSURE ~~\\ 
The learned semantic projections \textbf{A};
\STATE Denote $L^{(1)}  = AA^{'}  - M_I,L^{(2)}  = AB^{'} - M_C$
\FOR {t=1:T}
\STATE Decide the order of $m_1 \times q$ indices $(i,j)$ by random permutation $(i=1,...,m_1, j=1,...,q)$.
    \FOR {each of the $m_1 \times q$ indices $(i,j)$}
    \STATE Select the entry $A_{ij}$ to update;
    \STATE Compute $\partial g\left( {A_{i,j} } \right)$ and $\partial ^2 g\left( {A_{i,j} } \right)$ by
        \[
        \begin{array}{l}
        \partial g\left( {A_{i,j} } \right) = {4L_{i,*}^{(1)} A_{*, j} } + {2L_{i,*}^{(2)} B_{*, j} } \\
        \partial ^2 g\left( {A_{i,j} } \right) = {4\left( {A_{*, j}^{'} A_{*, j}  + A_{i,j}^2  + L_{i,i}^{(1)} } \right)} + {2B_{*, j}^{'} B_{*, j} } \\
        \end{array}
        \]
    \STATE Update $A_{i,j}  \leftarrow A_{i,j}  + d$, using
        $ d =  - \frac{{\partial g\left( {A_{i,j} } \right)}}{{\partial ^2 g\left( {A_{i,j} } \right)}}$
    \STATE Update Loss $L^{(1)}, L^{(2)}$ by
        \[
        \label{updateLoss}
        \begin{array}{l}
         L_{i,* }^{(1)}  \leftarrow L_{i, * }^{(1)}  + dA_{*, j}^{'} ,\mathop {}\nolimits_{} L_{*, i}^{(1)}  \leftarrow L_{*, i}^{(1)}  + dA_{*, j}  \\
         L_{i,i}^{(1)}  \leftarrow L_{i,i}^{(1)}  + d^2 ,\mathop {}\nolimits_{} L_{i,* }^{(2)}  \leftarrow L_{i,* }^{(2)}  + dB_{*, j}^{'}  \\
         \end{array}
        \]
    \ENDFOR
    \STATE Break, when
    \[
     {{abs\left( {Loss(t) - Loss(t - 1)} \right)} \mathord{\left/
        {\vphantom {{abs\left( {Loss(t) - Loss(t - 1)} \right)} {Loss(t - 1) \le \varepsilon }}} \right.
         \kern-\nulldelimiterspace} {Loss(t - 1) \le \varepsilon }}
    \]
\ENDFOR
\end{algorithmic}
\end{algorithm}

The time complexity of MPL-CD is ${\rm O}\left( {Tq{m^2}} \right)$ with small $T$ and $q$. $m$ represents $m_1$ in (\ref{subop1}), $m_2$ in (\ref{subop2}), which is also moderately small. Therefore, the whole process converges efficiently.

\begin{table}
\small
\centering
\caption{ Mean Average Precision (MAP) for \textbf{Wiki}: SIFT128+Topic10, in \%}
\begin{tabular}{|c|c|c|c|c|c|c|}
\hline
 \multirow{2}{*}{} & \multirow{2}{*}{} & \multicolumn{5}{|c|}{dimensions of shared space} \\
\cline{3-7}
 &  & 8 & 10 & 16 & 32 & 64\\
\cline{3-7}
\hline

\multirow{3}{*}{$Q_{I \to I}$}  & {SCM} & - & 16.0  & - & -  & -  \\
\cline{2-7}
 & {Marg-SM} & - & 17.3 & - & -  & -  \\
\cline{2-7}
\cline{2-7}
 & {LSSH} & 19.2 & 21.0  & 21.2 & 22.6  & 22.6  \\
\cline{2-7}
 & {CMSRM} & 13.1 & 12.4 & - & -  & -  \\
\cline{2-7}
 & {SDSRL} & \textbf{22.8} & \textbf{22.8}  & \textbf{22.8} & \textbf{22.8}  & \textbf{22.8}  \\
\hline

\multirow{3}{*}{$Q_{I \to T}$}  & {SCM} & - & 26.3  & - & -  & -  \\
\cline{2-7}
 & {Marg-SM} & - & \textbf{27.9} & - & -  & -  \\
\cline{2-7}
 & {BITR} & 14.5 & 13.6 & - & -  & -  \\
\cline{2-7}
 & {LSSH} & 19.2 & 21.0  & 21.2 & 22.6  & 22.6  \\
\cline{2-7}
 & {CMSRM} & 22.1 & 19.7 & - & -  & -  \\
\cline{2-7}
 & {SDSRL} & \textbf{26.5} & 26.8  & \textbf{26.8} & \textbf{26.8}  & \textbf{26.8}  \\
\hline

\multirow{3}{*}{$Q_{T \to I}$}  & SCM & - & 26.7 & - & -  & -  \\
\cline{2-7}
 & {Margin-SM} & - & 31.1 & - & -  & -  \\
\cline{2-7}
 & {BITR} & 14.1 & 11.5 & - & -  & -  \\
\cline{2-7}
 & LSSH & 42.9 & 47.7  & 49.5 & 51.9  & 52.8  \\
\cline{2-7}
 & {CMSRM} & 16.7 & 16.0 & - & -  & -  \\
\cline{2-7}
 & {SDSRL}& \textbf{59.8} & \textbf{63.2}  & \textbf{63.2} & \textbf{63.3}  & \textbf{63.2}  \\
\hline

\multirow{3}{*}{$Q_{T \to T}$}  & SCM & - & 59.5 & - & -  & -  \\
\cline{2-7}
 & {Margin-SM} & - & 61.0 & - & -  & -  \\
\cline{2-7}
\cline{2-7}
 & LSSH & 42.9 & 47.7  & 49.5 & 51.9  & 52.8  \\
\cline{2-7}
 & {CMSRM} & 46.1 & 47.9 & - & -  & -  \\
\cline{2-7}
 & {SDSRL}& \textbf{58.9} & \textbf{62.4}  & \textbf{62.3} & \textbf{61.5}  & \textbf{61.8}  \\
\hline
\end{tabular}
\label{map_wiki_xmodal}
\end{table}

\section{Experiments\label{experiments}}
We compare our proposed method, \textbf{SDSRL}, with several state-of-the-art methods, such as SCM \cite{Xmodal_TPAMI2014}, LSSH  \cite{ZhouJile2014_LSSH}, CMSRM \cite{Wufei2013}, BITR \cite{im2text_VermaJ14}, 
Marginal-SM \cite{menon-SDM2015}.
The experiments are conducted on two public real-world dataset.

In order to evaluate algorithms' robustness against different raw feature selections, we intensively perform cross-modal retrievals with different raw feature inputs. All input features are L2-normalized to unit length.

The performances are measured by:(1) \textbf{Mean Average Precision}; (2) \textbf{Precision Recall curves}.
The experiment results are reported both on inter-media and intra-media retrieval tasks:(1) $Q_{I \to T}$ ranking texts from image query;
(2) $Q_{T \to I} $ ranking images from text queries; (3) $Q_{I \to I}$ ranking images from image queries; and (4) $Q_{T \to T}$ ranking text from text queries.

The kernels used are RBF kernel $e^{ - \sigma \left\| {x_i  - x_j } \right\|^2 }$ with empirical $\sigma=1$ for both image and text data.
The num of outer loops is set as $E=50$ for \textbf{SDSRL}, and the inner iteration num as $T=10$ for MPL-CD.


\begin{table}
\small
\centering
\caption{Mean Average Precision (MAP) for \textbf{Wiki}: BoW1K+TFIDF5K, in \%}
\begin{tabular}{|c|c|c|c|c|c|c|}
\hline
 \multirow{2}{*}{} & \multirow{2}{*}{} & \multicolumn{5}{|c|}{dimensions of shared space} \\
\cline{3-7}
 &  & 8 & 10 & 16 & 32 & 64\\
\cline{3-7}
\hline

\multirow{3}{*}{$Q_{I \to I}$}  & {SCM} & - & 11.2  & - & -  & -  \\
\cline{2-7}
 & {Marg-SM} & - & 17.5 & - & -  & -  \\
\cline{2-7}
 & {LSSH} & 15.8 & 15.8 & 16.3 & 15.6  & 14.5  \\
\cline{2-7}
 & {CMSRM} & 12.6 & 12.7 & 12.9 & 12.3  & 12.5  \\
\cline{2-7}
 & {SDSRL} & \textbf{23.6} & \textbf{23.5}  & \textbf{23.5} & \textbf{23.5}  & \textbf{23.5}  \\
\hline

\multirow{3}{*}{$Q_{I \to T}$}  & {SCM} & - & 11.2  & - & -  & -  \\
\cline{2-7}
 & {Marg-SM} & - & 26.3 & - & -  & -  \\
\cline{2-7}
 & {BITR} & 13.3 & 13.3 & 12.8 & 12.4  & 12.0  \\
\cline{2-7}
 & {LSSH} & 15.8 & 15.8 & 16.3 & 15.6  & 14.5  \\
\cline{2-7}
 & {CMSRM} & 19.7 & 20.9 & 19.1 & 17.3  & 18.3  \\
\cline{2-7}
 & {SDSRL} & \textbf{30.4} & \textbf{30.1}  & \textbf{30.1} & \textbf{30.1}  & \textbf{30.1}  \\
\hline

\multirow{3}{*}{$Q_{T \to I}$}  & SCM & - & 11.2 & - & -  & -  \\
\cline{2-7}
 & {Marg-SM} & - & 31.4 & - & -  & -  \\
\cline{2-7}
 & {BITR} & 15.1 & 15.1 & 15.2 & 15.4  & 14.8  \\
\cline{2-7}
 & LSSH & 28.8 &  27.4  & 30.3 & 32.4  & 31.2  \\
\cline{2-7}
 & {CMSRM} & 17.3 & 19.6 & 19.5 & 16.3  & 19.0  \\
\cline{2-7}
 & {SDSRL}& \textbf{68.2 }& \textbf{70.3}  &\textbf{ 70.3} & \textbf{70.3}  &  \textbf{70.3} \\
\hline

\multirow{3}{*}{$Q_{T \to T}$}  & SCM & - & 11.2 & - & -  & -  \\
\cline{2-7}
 & {Marg-SM} & - & 55.1 & - & -  & -  \\
\cline{2-7}
\cline{2-7}
 & LSSH & 28.8 &  27.4  & 30.3 & 32.4  & 31.2  \\
\cline{2-7}
 & {CMSRM} & 51.7 & 47.9 & 52.1 & 52.4  & 52.0  \\
\cline{2-7}
 & {SDSRL}& \textbf{82.4} & \textbf{85.9} & \textbf{85.9} & \textbf{85.9} & \textbf{ 86.0} \\
\hline

\end{tabular}
\label{map_wiki_1k5k}
\end{table}

\subsection{Wiki dataset}
The dataset contains 2,866 image/text pairs belonging to 10 semantic categories.  The text length is about 200 words.
As did in LSSH\cite{ZhouJile2014_LSSH}, We randomly select 75\% of the dataset as database and the rest as the query set.
Documents are considered to be similar if they belong to the same category.

Comprehensive experiments are conducted in case of two widely accepted feature types:
(1)\emph{\textbf{Wiki}: SIFT128 + Topic10} is the same as the ones used in SCM\cite{Xmodal_TPAMI2014}\footnote{http://www.svcl.ucsd.edu/projects/crossmodal/} and
(2)\emph{\textbf{Wiki}: BOW1K + TFIDF5K} as in CMSRM\cite{Wufei2013}\footnote{https://luxinxin.info/research}.

As {SCM\cite{Xmodal_TPAMI2014}} computes posterior class probabilities as its final semantic features through traditional multi-class SVM,
the final dimension $q$ of {SCM\cite{Xmodal_TPAMI2014}} is therefore kept to be constant(q=10, the class num). The similar case is for {Marginal-SM\cite{menon-SDM2015}}. For both CMSRM and BITR learn embedded
space using either CCA or structural SVM, their objective dimensions $q$ subject to the constraints that $q\le 10$ for \emph{SIFT128 + Topic10}
and $q\le 1000$ for \emph{BOW1K + TFIDF5K}.

We empirically set the dimensions of feature mappings in kernel approximation stage: $m_1=1000$, $m_2=20$ for \emph{\textbf{Wiki}: SIFT128 + Topic10};
$m_1=1000$,$m_2=1000$ for \emph{ \textbf{Wiki}: BOW1K + TFIDF5K}.

\begin{table}
\small
\centering
\caption{Mean Average Precision (MAP) for \textbf{NUSWIDE}: BOW500+Tags1K, in \%}
\begin{tabular}{|c|c|c|c|c|c|c|}
\hline
 \multirow{2}{*}{} &  \multirow{2}{*}{} & \multicolumn{5}{|c|}{dimensions of shared space} \\
\cline{3-7}
 &  & 10 & 16 & 32 & 64 & 128 \\
\cline{3-7}
\hline
\multirow{3}{*}{$Q_{I \to I}$}
 & LSSH  & 42.4 & 41.9 & 42.0 & 41.6  & 41.2  \\
\cline{2-7}
 & {CMSRM}  & 39.6 & 43.3 & 39.0 & 44.7 & 39.1 \\
 \cline{2-7}
 & SDSRL  & \textbf{50.0} & \textbf{50.2} & \textbf{50.1} & \textbf{50.0}  & \textbf{50.0}  \\
\hline

\multirow{3}{*}{$Q_{I \to T}$} & BITR & 48.4 & 46.3 & 44.2 & 43.8 & 43.4  \\
\cline{2-7}
 & LSSH  & 42.4 & 41.9 & 42.0 & 41.6  & 41.2  \\
\cline{2-7}
 & {CMSRM}  & 46.9 & 50.2 & 49.8 & 51.5 & 45.9 \\
\cline{2-7}
 & SDSRL & \textbf{54.9} & \textbf{54.9} & \textbf{54.9} & \textbf{54.9} & \textbf{54.9} \\
\hline

\multirow{3}{*}{$Q_{T \to I}$} & BITR  & 49.4 & 49.0 & 48.6 & 48.3 & 0.471 \\
\cline{2-7}
 & LSSH & 45.4 & 44.4  & 44.8 & 44.3  & 43.2  \\
\cline{2-7}
 & {CMSRM}  & 44.9 & 49.0 & 48.6 & 49.5 & 43.6 \\
\cline{2-7}
 & SDSRL  & \textbf{52.9} & \textbf{52.8}  & \textbf{52.8} & \textbf{52.6}  & \textbf{52.6}  \\
\hline

\multirow{3}{*}{$Q_{T \to T}$}
 & LSSH  & 45.4 & 44.4  & 44.8 & 44.3  & 43.2  \\
\cline{2-7}
 & {CMSRM}  & 49.1 & 63.5 & 59.5 & 63.3 & 58.6 \\
\cline{2-7}
 & SDSRL  & \textbf{63.7} & \textbf{63.8} & \textbf{63.8} & \textbf{63.8}  & \textbf{63.8}  \\
\hline
\end{tabular}
\label{NUSWIDE:map_bow500}
\end{table}

\subsection{NUSWIDE}
The dataset was annotated by 81 concepts, but some are scarce. Similar to LSSH\cite{ZhouJile2014_LSSH}, we select the 10 most common concepts and 1000 most frequent tags,
ensuring that each selected image-tag pair contains at least one tag and one of the top 10 concepts. Thus 181,365 images are left from the 269,648 images.
We randomly select 2000 images and the corresponding tags features as the query set. The rest are treated as database.
Pairs are considered to be similar if they share at least one concept.

Two different types of features are experimented: \emph{NUSWIDE: BoW500 + Tags1K} \cite{nuswide-civr09} and \emph{NUSWIDE: VLAD128 + Tags1K}.
The difference is the former extract 500-dim BoW features and the later 128-dim VLAD features\cite{jegou:vlad128_eccv2012} for image.
For simplicity, we empirically set $m_1=1000$, $m_2=1000$ as dimensions of feature lifting maps in both cases.

\begin{figure}
  \centering
  \includegraphics[width=8cm, height=6cm]{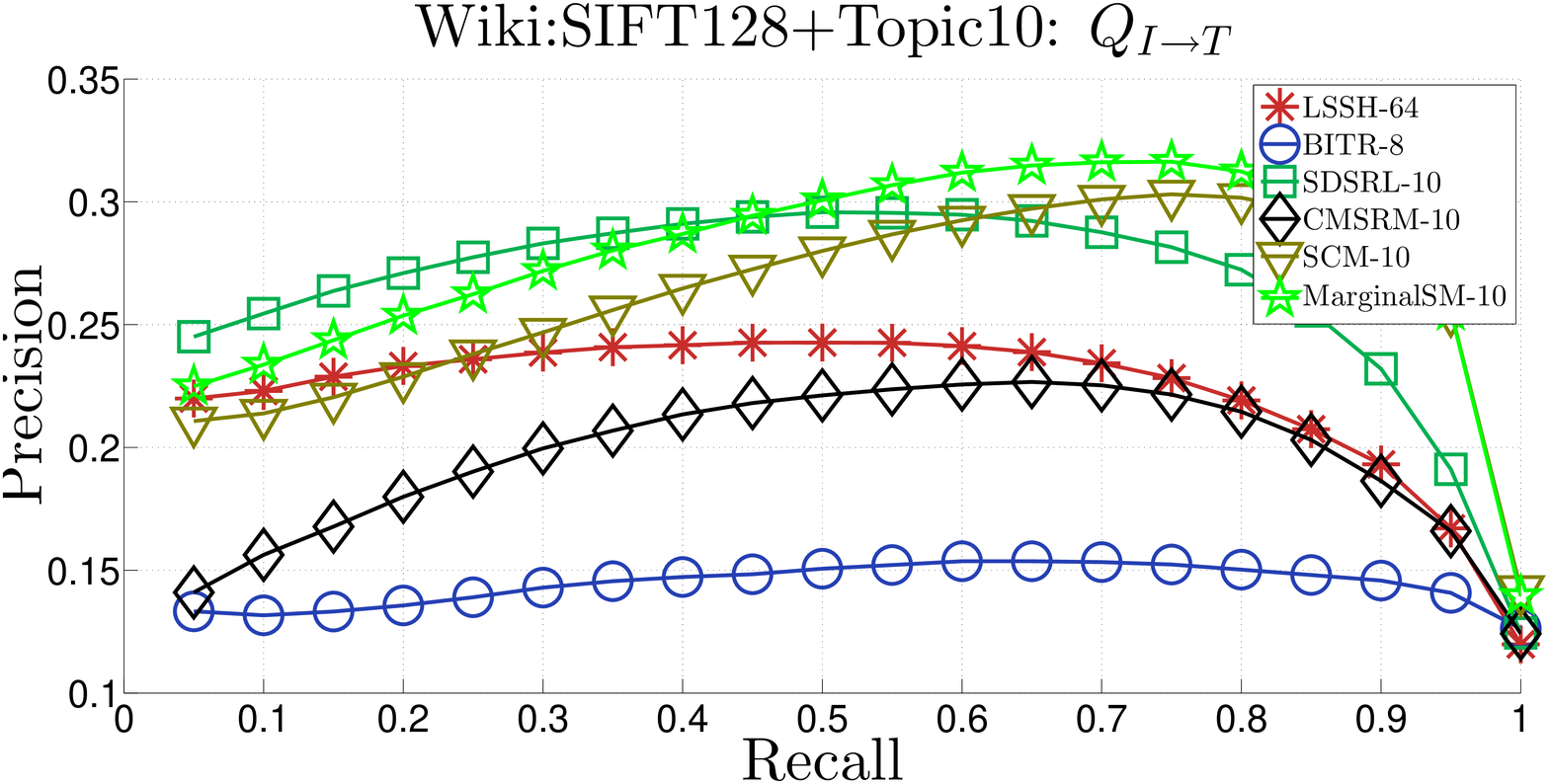}
	  
  \includegraphics[width=8cm, height=6cm]{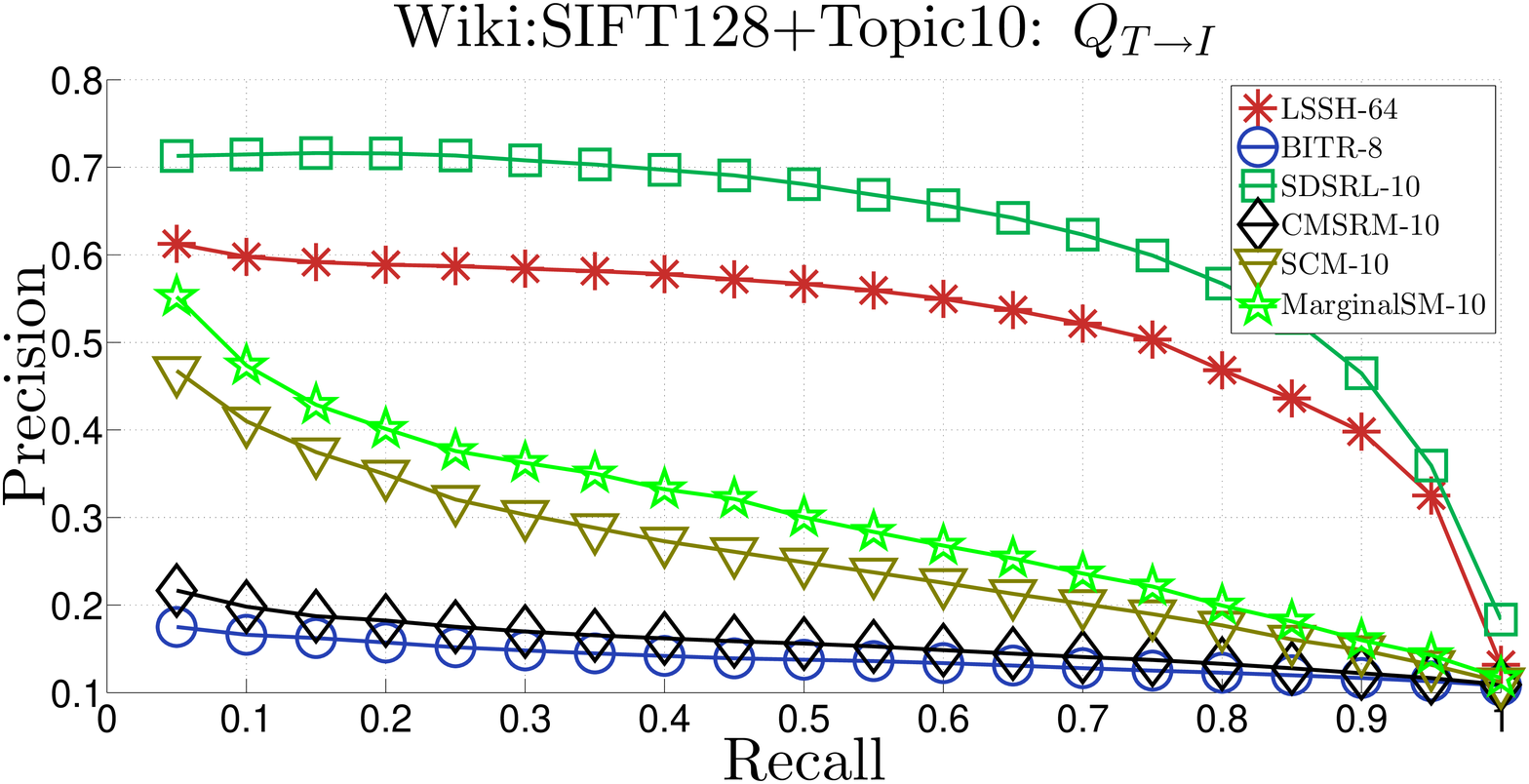}
	  
  \includegraphics[width=8cm, height=6cm]{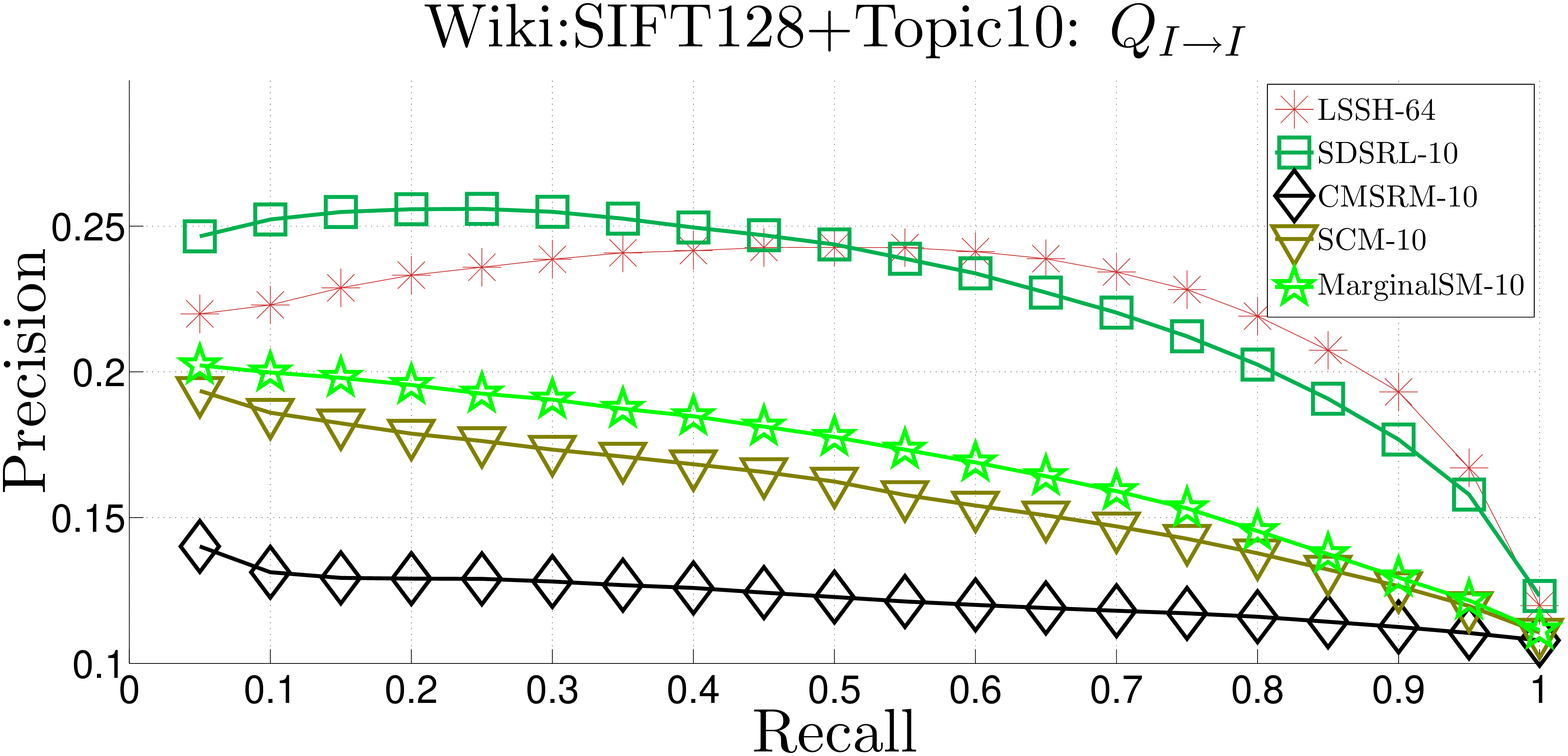}
	  
  \includegraphics[width=8cm, height=6cm]{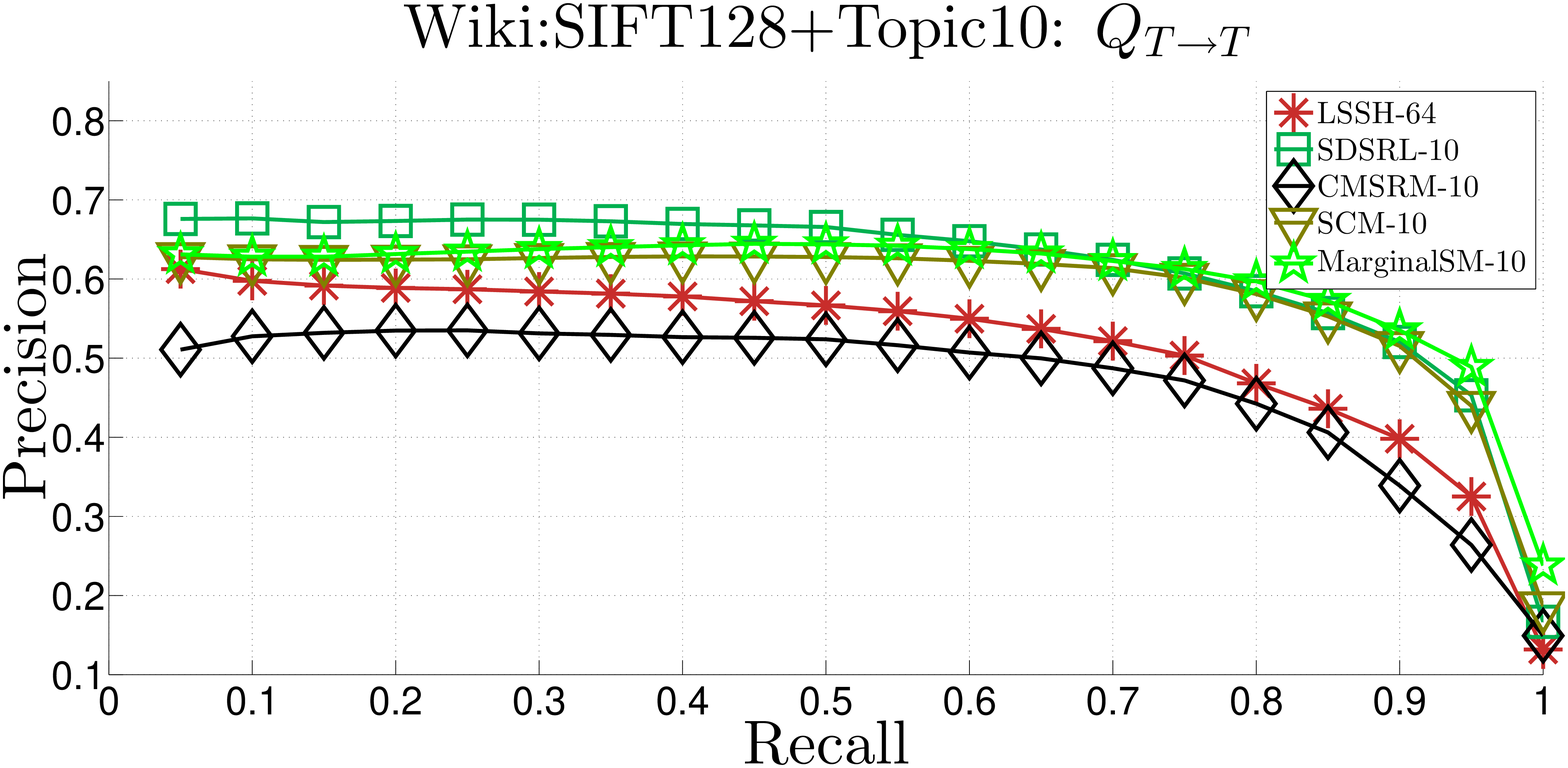}
  \caption{Precision-Recall Curve for Wiki: SIFT128+Topic10}
  \label{fig:wiki_xmodal_prc}
\end{figure}

\begin{figure}
    \center
        \includegraphics[width=8cm, height=6cm]{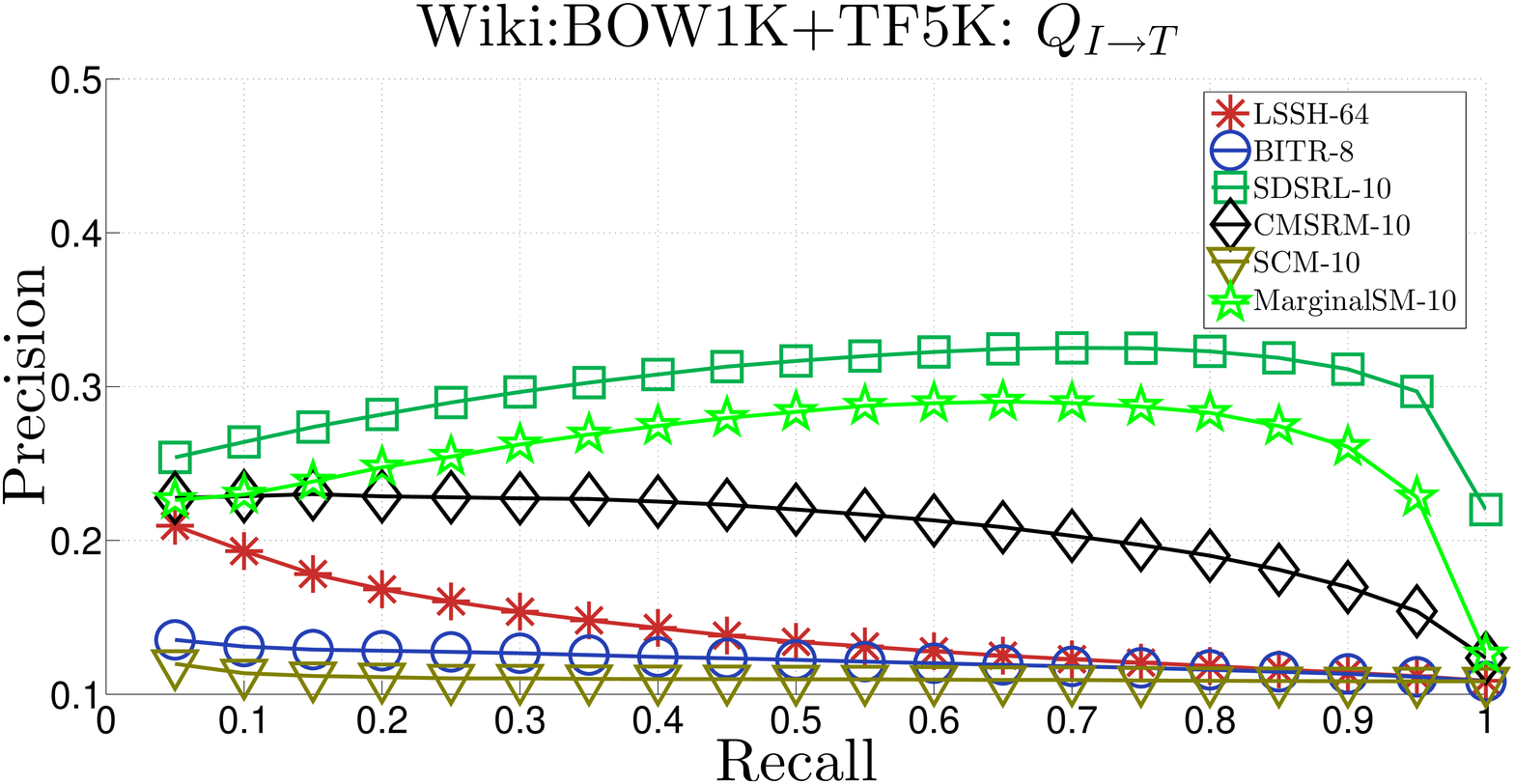}
       
        \includegraphics[width=8cm, height=6cm]{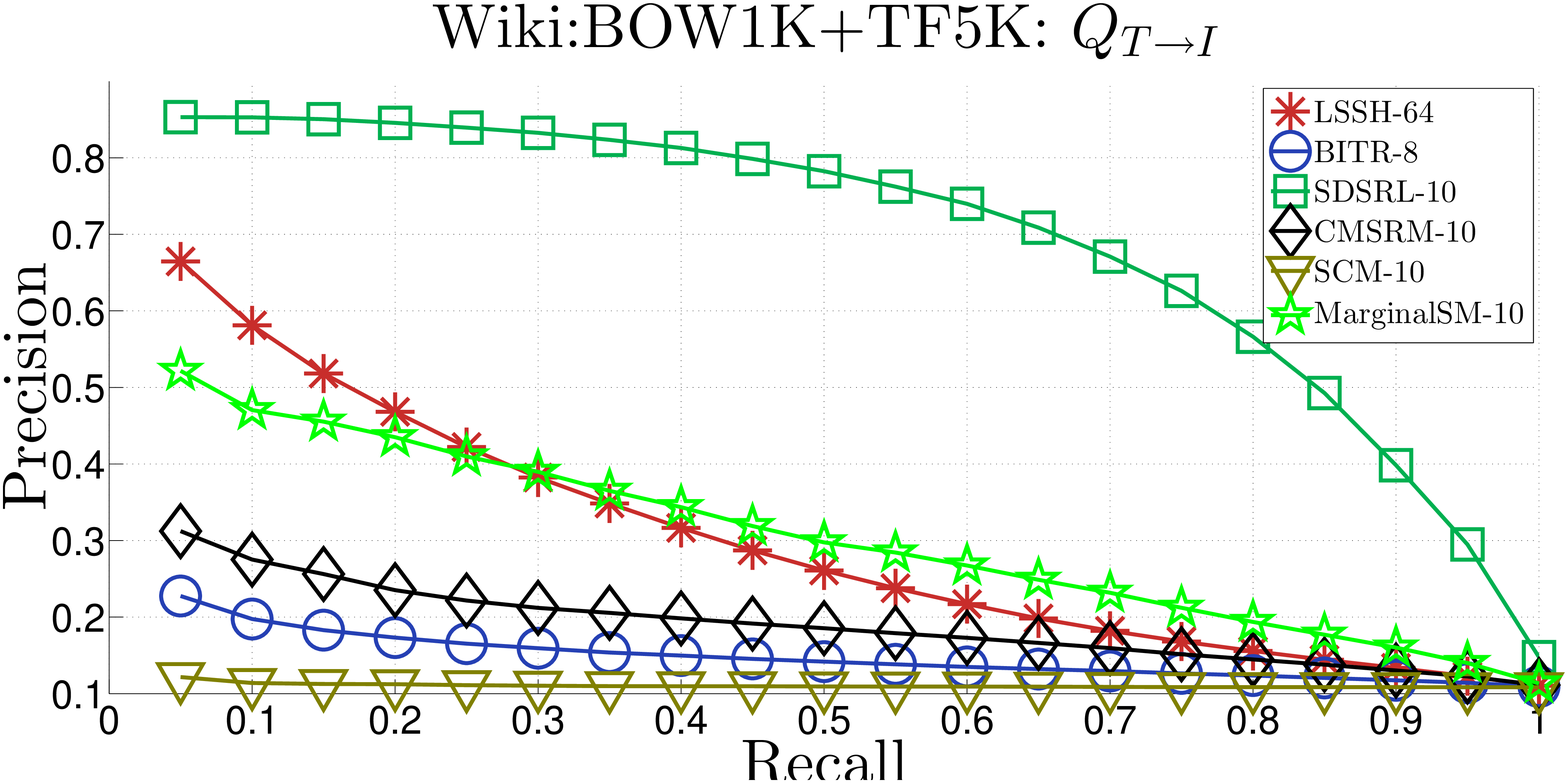}
        \includegraphics[width=8cm, height=6cm]{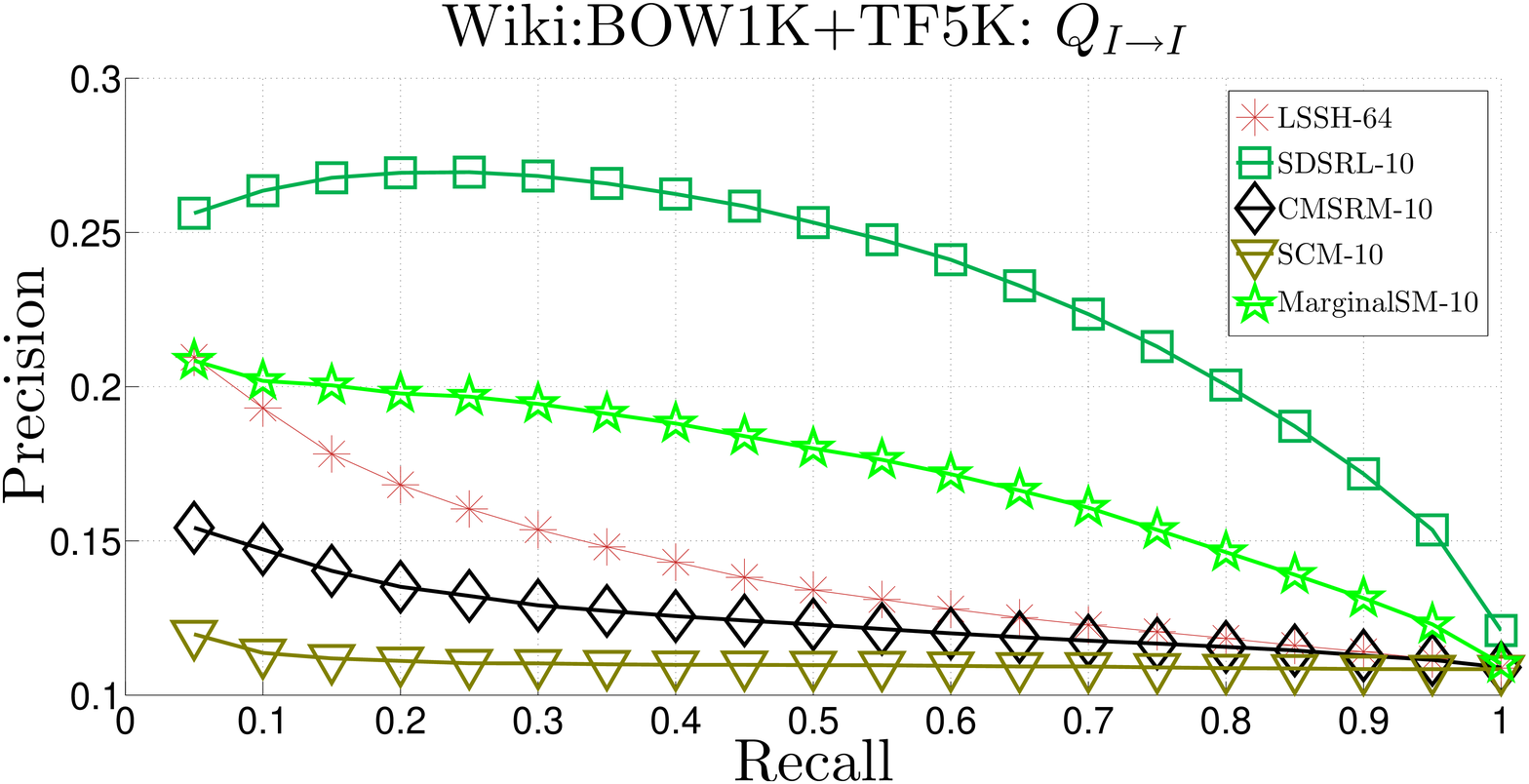}
        \includegraphics[width=8cm, height=6cm]{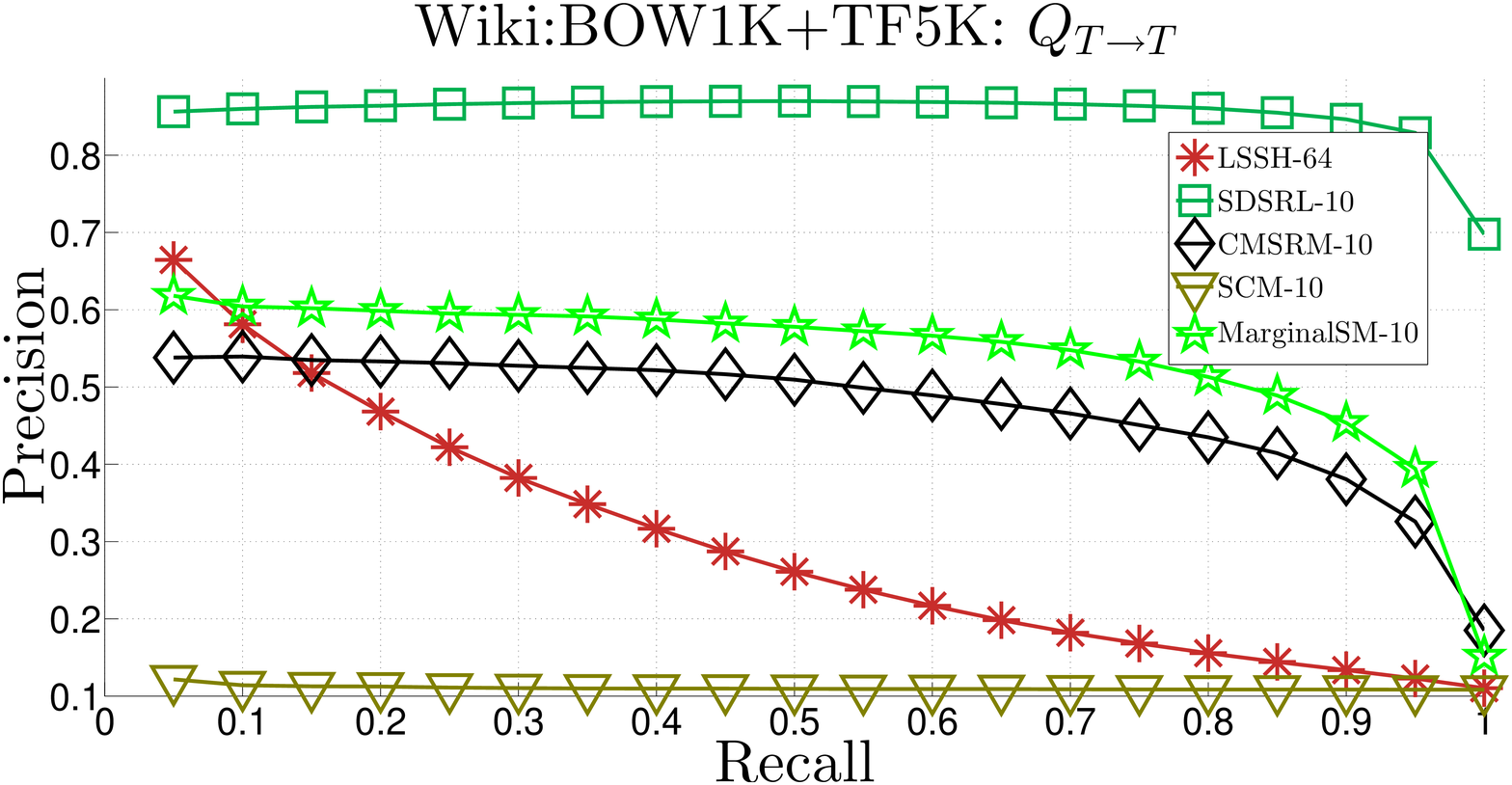}
    \caption{Precision-Recall Curve for Wiki: BOW1K+TFIDF5K}
    \label{fig:wiki_1k5k_prc}
\end{figure}

\begin{figure}
    \centering
        \includegraphics[width=8cm, height=6cm]{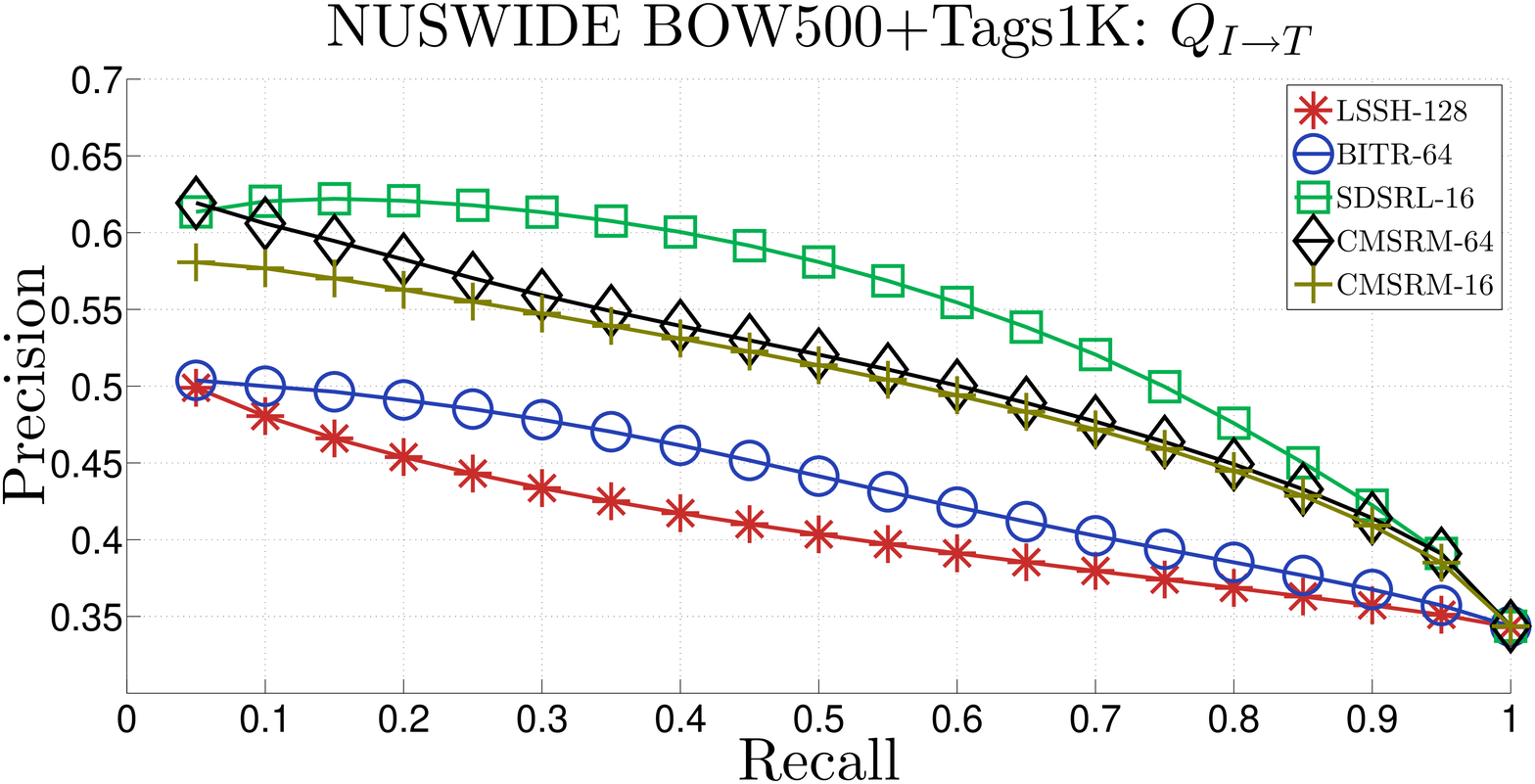}
        
        \includegraphics[width=8cm, height=6cm]{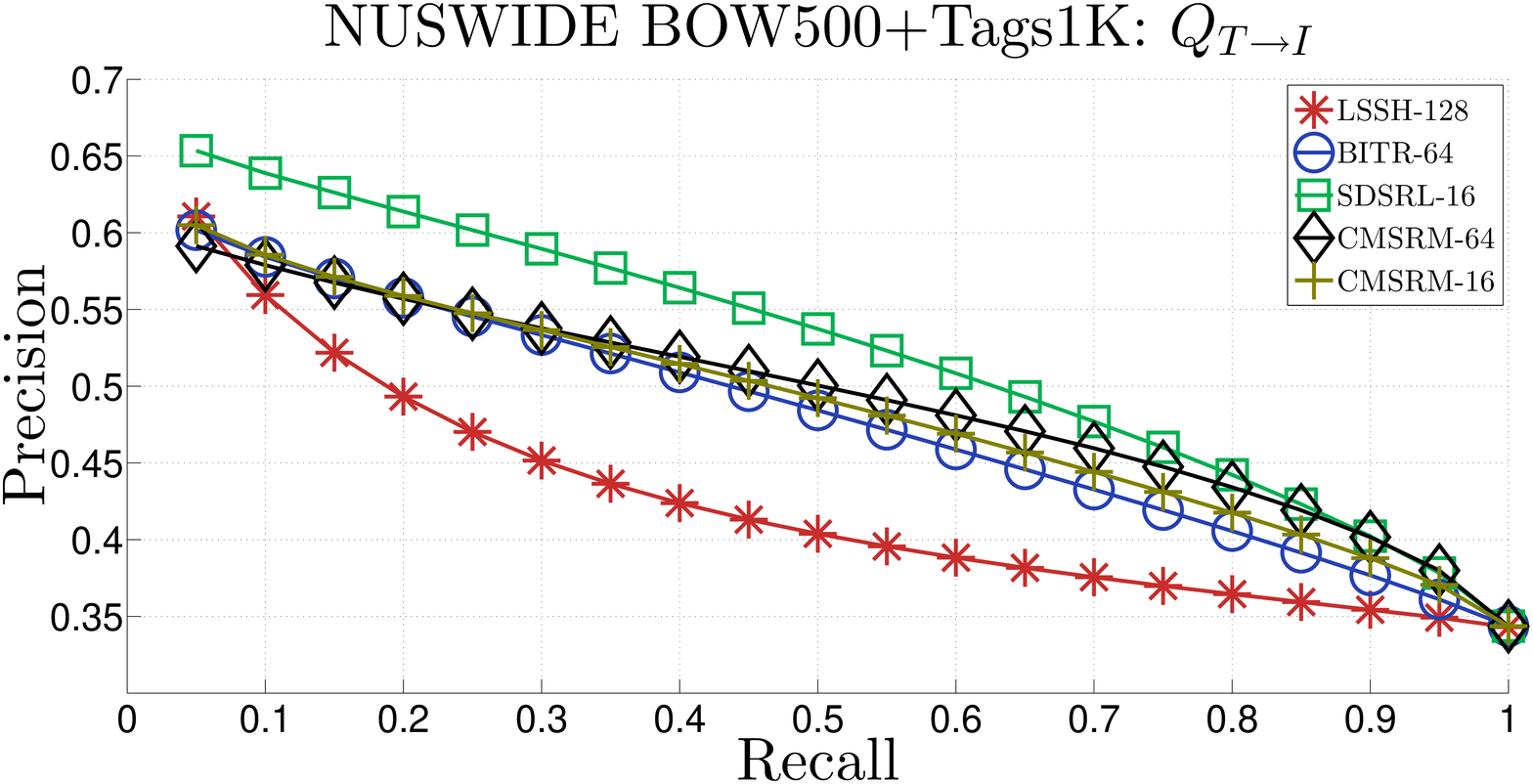}
        
        \includegraphics[width=8cm, height=6cm]{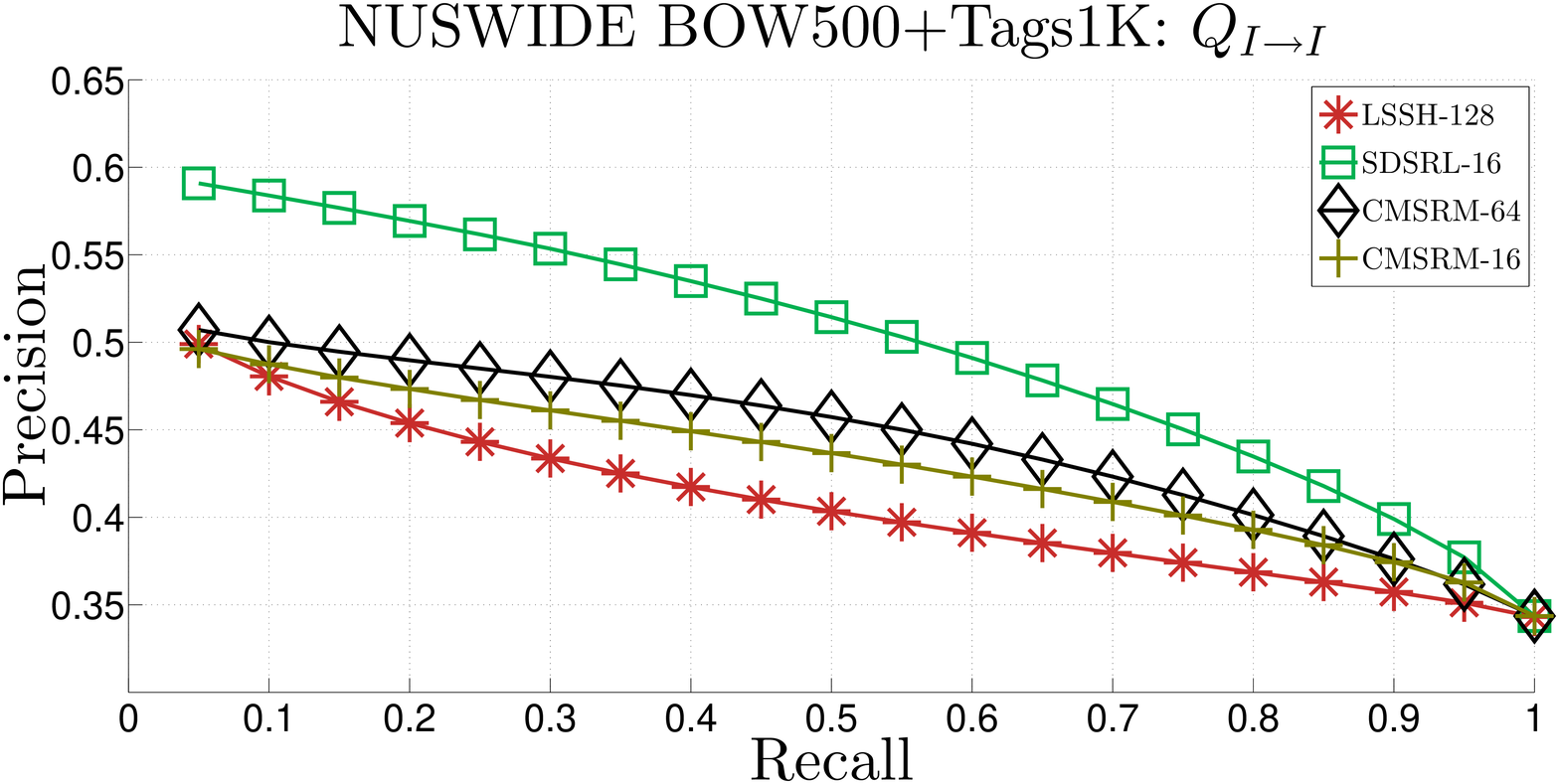}
        
        \includegraphics[width=8cm, height=6cm]{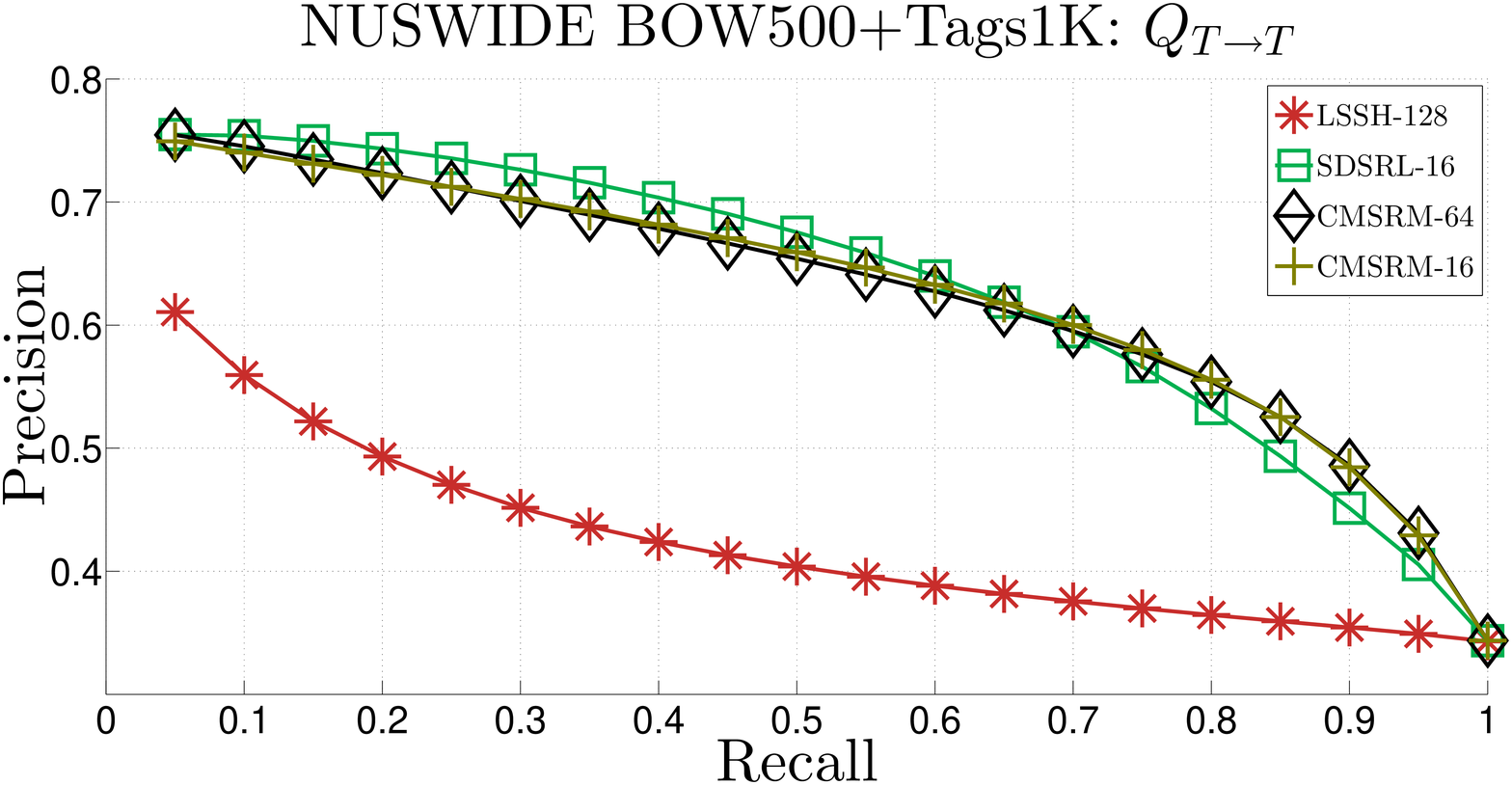}
        
    \caption{Precision-Recall Curve for NUSWIDE: BOW500+Tag1K}
    \label{fig:nuswide_bow500_prc}
\end{figure}

\begin{figure}
    \center
        \includegraphics[width=8cm, height=6cm]{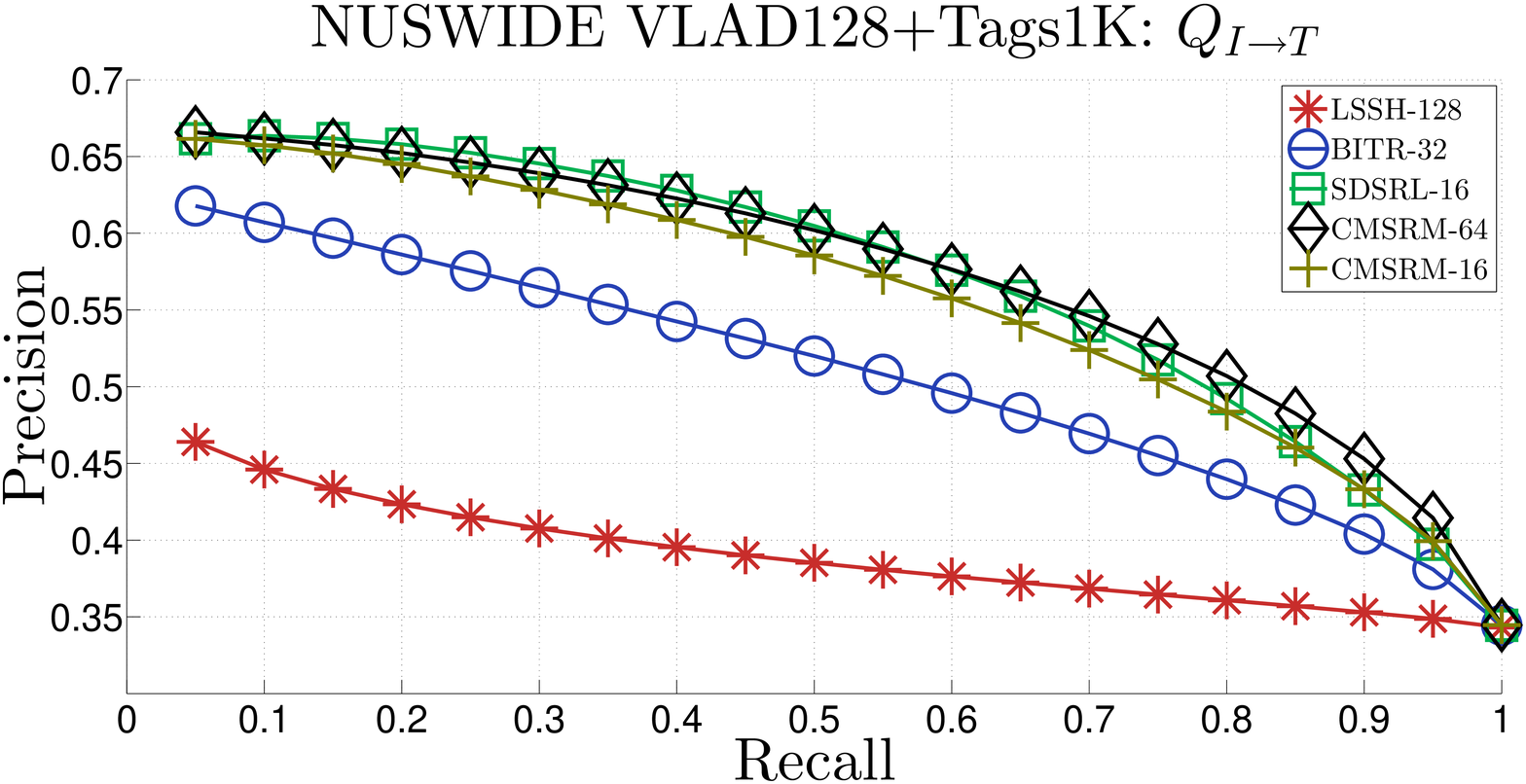}
        
        \includegraphics[width=8cm, height=6cm]{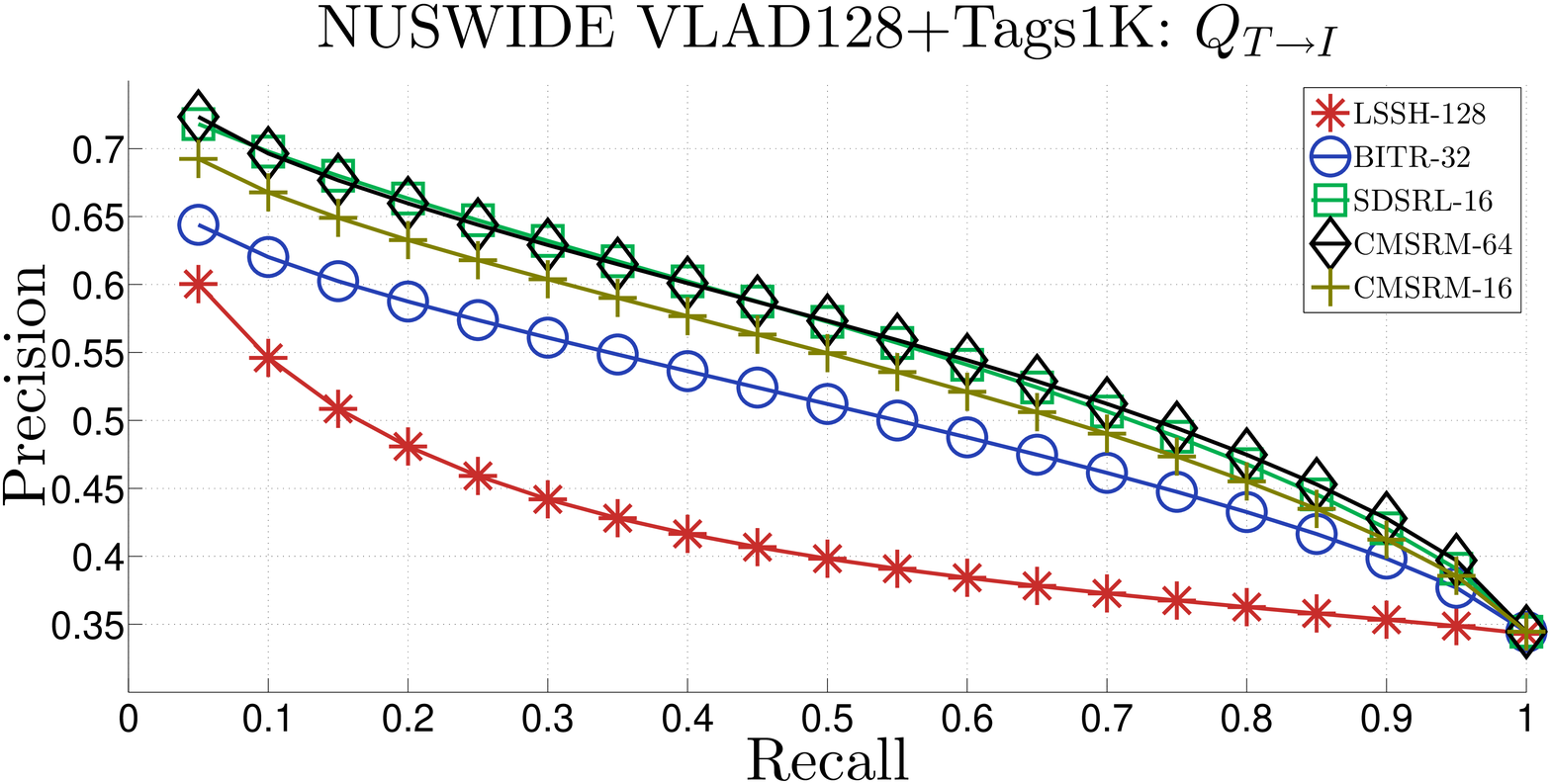}
        
        \includegraphics[width=8cm, height=6cm]{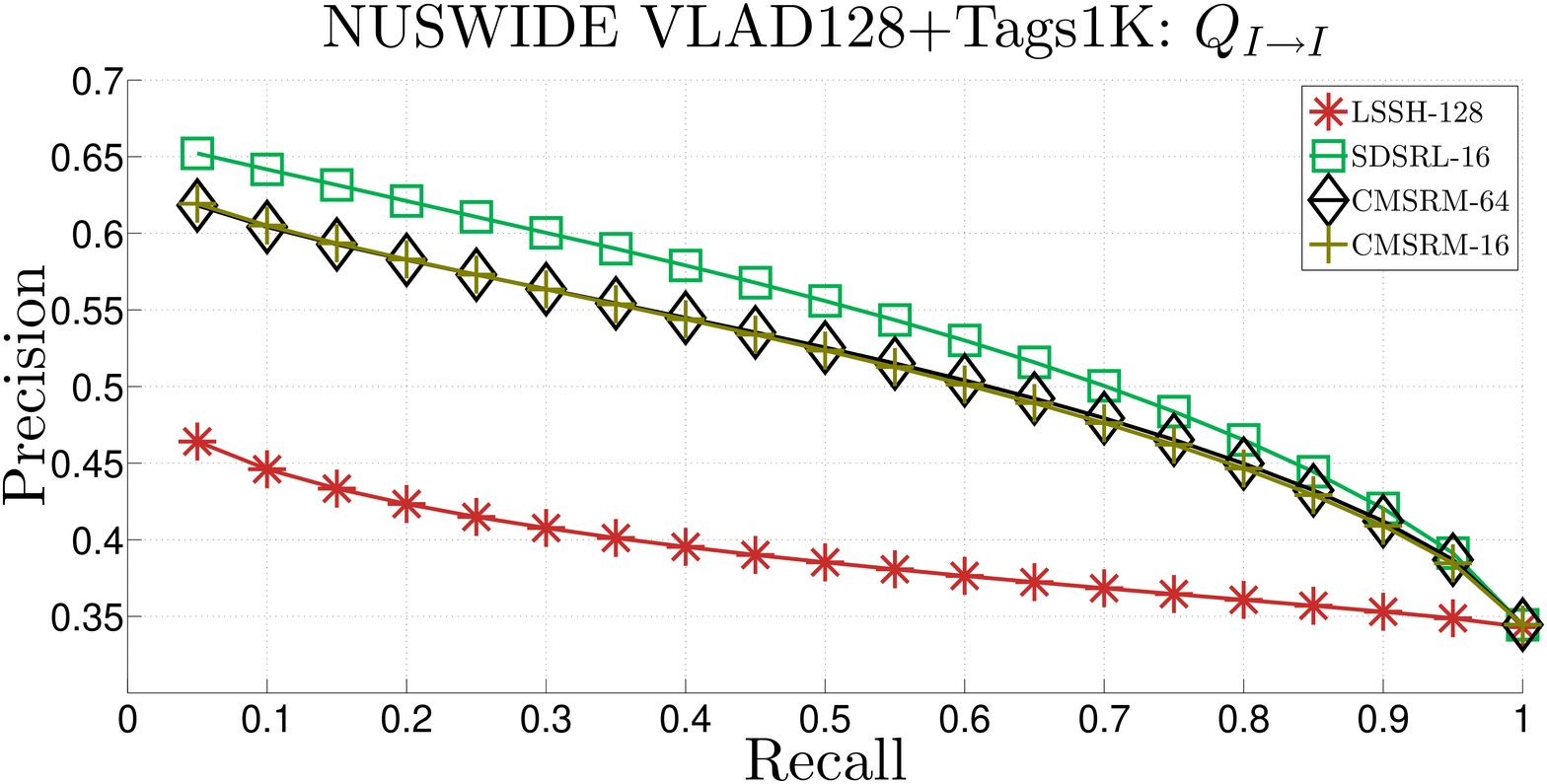}
        
        \includegraphics[width=8cm, height=6cm]{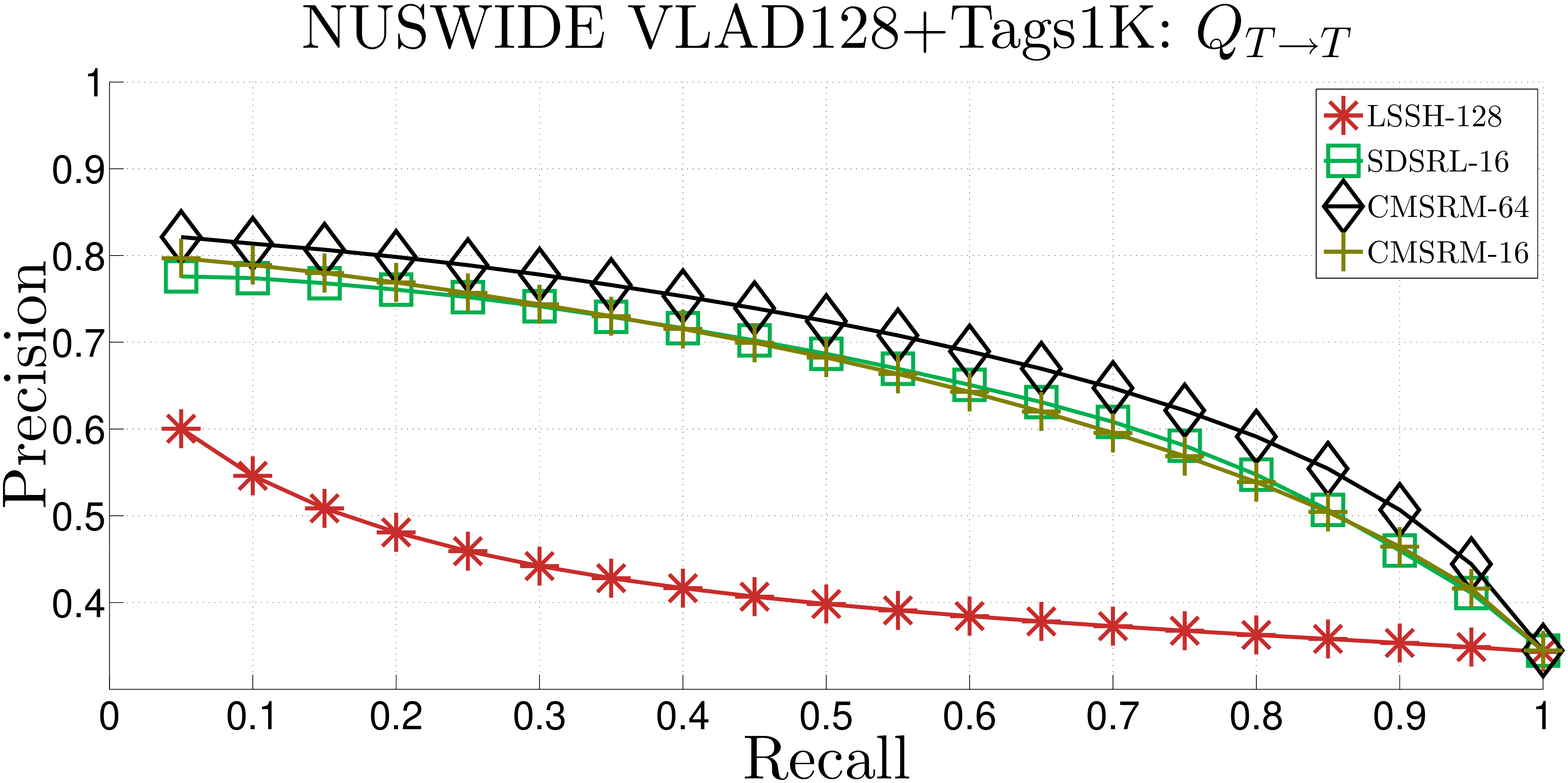}
        
    \caption{Precision-Recall Curve for NUSWIDE: VLAD128+Tag1K}
    \label{fig:nuswide_vlad128_prc}
\end{figure}

The MAP scores on the two dataset are reported in Tab.~\ref{map_wiki_xmodal}, Tab.~\ref{map_wiki_1k5k}, Tab.~\ref{NUSWIDE:map_bow500} and Tab.~\ref{NUSWIDE:map_vlad128}.
For more effecient comparison, we select repective best case from each compared method, and draw their precision-recall curves.
The curves are shown in Fig.~\ref{fig:wiki_xmodal_prc}, Fig.~\ref{fig:wiki_1k5k_prc}, Fig.~\ref{fig:nuswide_bow500_prc} and Fig.~\ref{fig:nuswide_vlad128_prc}.
In case of other selections, the trend of curves are generally similar. 
It should be noted that SCM\cite{Xmodal_TPAMI2014} and Marginal-SM\cite{menon-SDM2015}, proposed for multi-class data, can not be directly generalized to NUSWIDE, 
because NUSWIDE is in fact a multi-labeled dataset. Furthermore, BITR\cite{im2text_VermaJ14} aims to learn bilateral correlation matrix without the ability to obtain explicit multimodal features. 
Thus it can not access different modality data independently in its implicitly embedded space. So we also ignore the performance of BITR\cite{im2text_VermaJ14} on intra-modal retrieval tasks.

\begin{table}
\small
\centering
\caption{Mean Average Precision (MAP) for \textbf{NUSWIDE}: VLAD128+Tags1K, in \%}
\begin{tabular}{|c|c|c|c|c|c|c|}
\hline
 \multirow{2}{*}{} &  \multirow{2}{*}{} & \multicolumn{5}{|c|}{dimensions of shared space} \\
\cline{3-7}
 &  & 10 & 16 & 32 & 64 & 128 \\
\cline{3-7}
\hline
\multirow{3}{*}{$Q_{I \to I}$}
 & LSSH  & 40.8 & 40.4  & 40.1 & 39.9  & 39.3  \\
\cline{2-7}
 & {CMSRM}  & 48.1 & 51.5 & 51.5 & 51.6  & 51.8  \\
 \cline{2-7}
 & SDSRL  & \textbf{54.3} & \textbf{54.3} & \textbf{54.3} & \textbf{54.3}  & \textbf{54.3} \\
\hline

\multirow{3}{*}{$Q_{I \to T}$} & BITR & 51.9 & 51.9 & 51.2 & 49.9 & 46.7 \\
\cline{2-7}
 & LSSH  & 40.8 & 40.4  & 40.1 & 39.9  & 39.3  \\
\cline{2-7}
 & {CMSRM}  & 54.7 & 56.4 & 56.3 & \textbf{57.8}  & 57.1  \\
\cline{2-7}
 & SDSRL  & \textbf{57.4} & \textbf{57.5}  & \textbf{57.4} & 57.4 & \textbf{57.4} \\
\hline

\multirow{3}{*}{$Q_{T \to I}$} & BITR  & 51.3 & 51.5  & 51.1 & 50.2 & 48.1 \\
\cline{2-7}
 & LSSH & 44.4 & 44.0  & 44.0 & 43.5  & 42.6  \\
\cline{2-7}
 & {CMSRM}  & 52.6 & 54.5 & 54.0 & \textbf{56.8}  & 55.4  \\
\cline{2-7}
 & SDSRL  & \textbf{56.5} & \textbf{56.5}  & \textbf{56.5} & 56.4  & \textbf{56.4} \\
\hline

\multirow{3}{*}{$Q_{T \to T}$}
 & LSSH  & 44.4 & 44.0  & 44.0 & 43.5  & 42.6  \\
\cline{2-7}
 & {CMSRM}  & 62.3 & \textbf{65.2} & 64.6 & \textbf{69.1}  & \textbf{66.1}  \\
\cline{2-7}
 & SDSRL  & \textbf{65.0} & \textbf{65.2}  & \textbf{65.2} & 65.2  & 65.2  \\
\hline
\end{tabular}
\label{NUSWIDE:map_vlad128}
\end{table}

According to the experiment results, we can observe that the proposed \textbf{SDSRL} outperforms the other state-of-the-art methods in most scenarios.
In details, from the MAP scores, the performance of \textbf{SDSRL} is consistently stable and becomes comparatively steady after a certain semantic dimension.
Interestingly, the dimension is coincident with the num of data's semantic category.
It demonstrate that the proposed \textbf{SDSRL} can learn the intrinsic manifolds of these multimodal data.
Moreover, the proposed method is more robust to different input cases.
As indicated by the experimental results, especially on Wiki dataset, except our method, the performances of other methods are more or less affected.
From the Precision-Recall Curves, our proposed method achieves almost the best performance for intra/inter modal retrieval tasks with features of the lowest dimension.

\emph{Discussions:} We owe the excellent performance of \textbf{SDSRL} to its flexible learning framework.
Compared with LSSH, \textbf{SDSRL} learns directly from more general semantic correlations, avoiding the one-to-one strictly paired constraints.
Compared with BITR, \textbf{SDSRL} can access different modality contents more conveniently through multimodal semantic projection.
Compared with CMSRM, the multistage architecture of \textbf{SDSRL} simplifies the representations learning process with less parameters to be learned.

In terms of speed and scalability, the speed depends on the method used in the kernel approximation, which in practical scales well and runs fast. 
Take NUSWIDE dataset for example. All the experiments run on i7-2.4GHz CPU. For about 180,000 training images whose features are 500-dim BOW, it takes 9min to extract approximation kernel features. 
It should be noted that the kernel approximation method is not limited to Nystroem method currently used. 
In the stage of semantic down projection, the bottleneck of the scalability lies in the computation for semantic correlation matrix. 
The optimization complexity is $O(Tqm^2)$ for our method. In the experiments, $q$ is the dimension of compact semantic feature, which is small (less than 64). 
$m$ is the dimension of lifted features using kernel approximation (1000 in this paper). 
Compared to data size, it is also reasonably small. $T$ is the iteration number (50). 
The learning time of this stage takes about 4min, which is independent of dataset size. 
In testing, only linear computations are involved both in feature up-lifting and down-projecting stages, which make it very fast.

We also note that the advantage of SDSRL is clear for the Wiki corpus but maybe less so for the NUSWIDE corpus. However, It instead varifies a conclusion that the Wiki corpus is more suitable for cross-modal retrieval research. 
Compared with Wiki, NUSWIDE corpus has less text information with only several keyword labels, which makes it more like a multi-label dataset. 

\section{Conclusion\label{conclusion}}
We have proposed an effective multimodal representations learning strategy for crossmodal retrieval.
Intra- and Inter-modal semantic correlations are structurally considered in a unified framework.
Different modality data can be conveniently projected into a shared, metric-comparable semantic space.
The experiments have demonstrated that in the learned space, for multimodal data, the proposed method can not only do inter-media retrieval with high performance,
but also retain its advantages in intra-media task. The proposed strategy \textbf{SDSRL} is very straightforward and robust to different raw feature selections.
We are confident that it will shed some light on future multimodal representation learning research.


%

\section*{The Details of Multimdal Coordinate Decent Solutions for SDSRL}
 Taking subproblem (\ref{subop1}) for example, we randomly chooses one entry $A_{(i,j)}$ in \textbf{A} to update while keeping other entries fixed. Let $A  = [A_{(*, 1)} ,A_{(*, 2)} ,...,A_{(*, m_1 )} ]$, where $A_{(*, j)}$ is the $j^{th}$ column in \textbf{A}.
 The objective in (\ref{subop1}) is rewritten as
 \begin{displaymath}
 \begin{array}{l}
 \mathop {\min }\limits_{A } \left\| {A_{*, j} A_{*, j}^{'}  - R^{(1)} } \right\| + \left\| {A_{*, j} B_{*, j}^{'}  - R^{(2)} } \right\| \\
  \end{array}
 \end{displaymath}
 where
 \begin{displaymath}
 R^{(1)}  = {M_I  - \sum\limits_{k \ne j} {A_{*, k} A_{*, k}^T } }, ~~
 R^{(2)}  = {M_C  - \sum\limits_{k \ne j} {A_{*, k} B_{*, k}^T } }
 \end{displaymath}

 By fixing the other entries in \textbf{A}, the objective w.r.t. $A_{ij}$ can be further rewritten as:
 \begin{tiny}
 \begin{displaymath}
 \begin{array}{l}
  g\left( {A_{i,j} } \right) = \sum\limits_{l = 1}^{m_1 } {\sum\limits_{k = 1}^{m_1 } {\left( {A_{l,j} A_{k,j}  - R_{l,k}^{(1)} } \right)^2 } }  +  \sum\limits_{t = 1}^{m_2 } {\sum\limits_{p = 1}^{m_1 } {\left( {A_{p,j} B_{t,j}  - R_{p,t}^{(2)} } \right)^2 } }  \\
   \propto \left( {\left( {A_{i,j}^2  - R_{i,i}^{(1)} } \right)^2  + 2\sum\limits_{k \ne i} {\left( {A_{i,j} A_{k,j}  - R_{i,k}^{(1)} } \right)^2 } } \right) + \sum\limits_{t = 1}^{m_2 } {\left( {A_{i,j} B_{t,j}  - R_{i,t}^{(2)} } \right)^2 }  \\
  \end{array}
 \end{displaymath}
 \end{tiny}
 Suppose we update $A_{i,j} \to A_{i,j}+d$, we approximate $g\left( {A_{i,j}  + d} \right)$ by a quadratic function via Taylor expansion:
 \begin{displaymath}
 \label{taylor}
 g\left( {A_{i,j}  + d} \right) \approx g(A_{i,j} ) + \partial g\left( {A_{i,j} } \right)d + \partial ^2 g\left( {A_{i,j} } \right)d^2
 \end{displaymath}
 where
 \begin{tiny}
 \begin{displaymath}
 \begin{array}{l}
  \partial g\left( {A_{i,j} } \right) = {4\sum\limits_{k = 1}^{m_1 } {\left( {A_{i,j} A_{k,j}  - R_{i,k}^{(1)} } \right)A_{k,j} } }
  + {2\sum\limits_{t = 1}^{m_2 } {\left( {A_{i,j} B_{t,j}  - R_{i,t}^{(2)} } \right)B_{t,j} } } \\
  \partial ^2 g\left( {A_{i,j} } \right) = {12A_{i,j}^2  - 4R_{i,i}^{(1)}  + 4\sum\limits_{k \ne i} {A_{k,j}^2 } }
  + {2\sum\limits_{t = 1}^{m_2 } {B_{t,j}^2 } } \\
  \end{array}
 \end{displaymath}
 \end{tiny}
 The newton direction of earlier defined Taylor expansion is $ d =  - \frac{{\partial g\left( {A_{i,j} } \right)}}{{\partial ^2 g\left( {A_{i,j} } \right)}}$.

 In order to accelerate the computation, we maintain the loss $L^{(1)}  = A A^{\prime}  - M_I, L^{(2)}  = AB^{\prime} - M_C$,
 then the first (second) order derivatives can be calculated as
 \begin{displaymath}
 \label{partialcomputing}
 \begin{array}{l}
  \partial g\left( {A_{i,j} } \right) = {4L_{i, * }^{(1)} A_{*, j} } + {2L_{i, * }^{(2)} B_{*, j} } \\
  \partial ^2 g\left( {A_{i,j} } \right) = {4\left( {A_{*, j}^{\prime} A_{*, j}  + A_{i,j}^2  + L_{i,i}^{(1)} } \right)} + {2B_{*, j}^{'} B_{*, j} } \\
  \end{array}
 \end{displaymath}

 Given $A_{i,j}  \leftarrow A_{i,j}  + d$, we update the loss as
 \begin{displaymath}
 \label{updateLoss}
 \begin{array}{l}
  L_{i,* }^{(1)}  \leftarrow L_{i,* }^{(1)}  + dA_{*, j}^{'} ,\mathop {}\nolimits_{} L_{*, i}^{(1)}  \leftarrow L_{*, i}^{(1)}  + dA_{*, j}  \\
  L_{i,i}^{(1)}  \leftarrow L_{i,i}^{(1)}  + d^2 ,\mathop {}\nolimits_{} L_{i,* }^{(2)}  \leftarrow L_{i,* }^{(2)}  + dB_{*, j}^{\prime}  \\
  \end{array}
 \end{displaymath}
 We will obtain the optimal $A$ by minimizing overall loss $Loss  = \left\| {L^{(1)}  } \right\| + \left\| {L^{(2)}  } \right\|$.



\bibliographystyle{unsrt}
\bibliography{MyBib}

\end{document}